\documentclass[aps,prb,twocolumn,superscriptaddress,amsmath]{revtex4}

\usepackage[dvips]{graphicx}
\usepackage{dcolumn}
\usepackage{amsmath}
\usepackage{bm}
\usepackage{verbatim}

\newcommand{\be}{\begin{equation}}
\newcommand{\ee}{\end{equation}}
\newcommand{\bea}{\begin{eqnarray}}
\newcommand{\eea}{\end{eqnarray}}
\newcommand{\bean}{\begin{eqnarray*}}
\newcommand{\eean}{\end{eqnarray*}}

\newcommand{\abs}[1]{\left| #1 \right|}

\newcommand{\htwoo}{H$_2$O}

\begin{document}

\widetext

\title{Quantum Monte Carlo study of the Protonated Water Dimer} 

\author{Mario Dagrada}
\email{mario.dagrada@impmc.upmc.fr}
\affiliation{CNRS and Institut de Min\'eralogie et 
de Physique des Milieux condens\'es,
Universit\'e Pierre et Marie Curie,
case 115, 4 place Jussieu, 75252, Paris cedex 05, France}
\author{Michele Casula}
\email{michele.casula@impmc.upmc.fr}
\affiliation{CNRS and Institut de Min\'eralogie et 
de Physique des Milieux condens\'es,
Universit\'e Pierre et Marie Curie,
case 115, 4 place Jussieu, 75252, Paris cedex 05, France}
\author{Antonino M. Saitta}
\email{marco.saitta@impmc.jussieu.fr}
\affiliation{CNRS and Institut de Min\'eralogie et
de Physique des Milieux condens\'es,
Universit\'e Pierre et Marie Curie,
case 115, 4 place Jussieu, 75252, Paris cedex 05, France}
\author{Sandro Sorella}
\email{sorella@sissa.it}
\affiliation{International School for Advanced Studies (SISSA) Via Beirut 2,4
  34014 Trieste, Italy and INFM Democritos National Simulation Center,
  Trieste, Italy} 
\author{Francesco Mauri}
\email{francesco.mauri@impmc.jussieu.fr}
\affiliation{CNRS and Institut de Min\'eralogie et 
de Physique des Milieux condens\'es,
Universit\'e Pierre et Marie Curie,
case 115, 4 place Jussieu, 75252, Paris cedex 05, France}

\date{\today}

\begin{abstract}
We report an extensive theoretical study of the protonated water dimer
$H_5O_2^+$ (Zundel ion) by means of the highly correlated
variational Monte Carlo and lattice regularized Monte Carlo
approaches. This system represents the simplest 
model for proton transfer (PT) and a correct description of its
properties is essential in order to understand the PT mechanism
in more complex acqueous systems.
Our Jastrow correlated AGP wave function ensures an 
accurate treatment of electron correlations. 
Exploiting the advantages of contracting the primitive basis 
set over atomic hybrid orbitals, we
are able to limit dramatically the number of variational parameters with a
systematic control on the numerical precision, a crucial ingredient in order to simulate larger systems. 
We investigate the energetics and geometrical properties of 
the $H_5O_2^+$ as a function of the oxygen-oxygen distance, taken as reaction coordinate.
In both cases, our QMC results are found in excellent agreement with 
state-of-the-art coupled cluster CCSD(T) techniques. Calculations on proton 
transfer static barriers and dissociation energies display the same 
agreement. A comparison with density functional theory results in the
PBE approximation points out the 
crucial role of electron correlations for a correct description 
of the PT in the dimer. 
Finally, the ability of our method to resolve very tiny 
energy differences ($\sim 0.1$ Kcal/mol) at which the proton hopping
takes place and the corresponding structural variations optimized
directly in the VMC framework is also proven. Our approach combines these 
features with a $N^3$-$N^4$ scaling with number of particles.  
This value is favorable with respect to other highly
correlated \emph{ab initio}
approaches and it allows the simulation of
more realistic PT models; a test calculation
on a larger protonated water cluster is hence carried out. 
The QMC approach used here represents a promising 
candidate to provide the first high-level \emph{ab-initio} description 
of PT in water. 

\end{abstract}

\maketitle

\section{Introduction}
\label{introduction}
Water is a key element for life, but nevertheless its properties
arising from its unique structure are not fully understood yet.  
A large amount of published works explored the several phases of water
and many efforts have been spent by the scientific 
community in order to match experimental findings with theoretical
predictions.

From the chemical and biological points of view, a deep understanding of
the properties of acqueous systems and therefore of liquid water is
fundamental. Over the past years,  
the structural and dynamical properties of liquid water have been
investigated relying on the state-of-the-art molecular dynamics
techniques such as  
empirical force field methods
\cite{force_field1,force_field2,force_field3} and \emph{ab initio}
molecular dynamics (AIMD).  
The latter methodology has been extensively employed for the study of
liquid water \cite{water1_parri, water2, water3, water4, water5} as
well as acqueous solution of  
biological interest \cite{abinitio_bio1,abinitio_bio2}; within,
this technique, the atomic force fields required as input by any
molecular dynamics simulation are constructed from \emph{ab initio}
potential energy surfaces (PESs) and  
therefore directly derived from first principles electronic structure
calculations.

Working out an accurate PES for liquid water has been a longstanding
challenge in the scientific community; 
this is due mainly to the absence of an \emph{ab initio} method able
to develop a reliable PES that describes breaking and formation of
hydrogen  
bonds; other issues are related to difficulties in simulating weak
interaction and polarization effects in the network of polar
molecules.

The simulation of liquid systems requires model clusters with a large
amount of molecules and therefore a good scalability is fundamental in
order to carry out simulations  
in a reasonable
computational time; therefore the technique which has been mostly
employed over the past years is density functional theory (DFT) which
combines scalability and flexibility 
features.
Nevertheless, molecular dynamics simulations which start from
DFT-based PES fail in reproducing basic properties of liquid water
such as the melting point 
temperature \cite{water_temperature} and the oxygen-oxygen radial
distribution function $g_{OO}$ \cite{water_radial}.

Another major problem which has not been 
elucidated yet by AIMD is the description of 
proton transfer (PT) in liquid water, phenomenon which goes under the
name  
of Grotthuss mechanism \cite{grotthuss,agmon,200years_after}. 
This process is extremely important in a wide range of biological
systems and it influences many dynamical processes in material
science, biochemistry  
and bioenergetics; for example, PT is one of the main mechanism for
charge transport through the cell membranes  
\cite{phys_rev, pnas_membrane}; it plays an important role in several
enzymathic reactions and photosynthesis \cite{photos_ref1,
photos_ref2} and moreover proton has the  
role of mediator and promoter of many 
acid-base reactions in solution
\cite{PTacqueous_sol1,PTacqueous_sol2,PTacqueous_sol3}.

Although over the past years many experimental and theoretical works
have been devoted to the study of Grotthuss reaction, an accurate
quantitative explanation  
of the underlined mechanism is not fully reached yet.
Theoretical investigations still fail in PT basic properties predictions, 
such as the activation barrier. Experimental measures of this quantity 
have been carried out \cite{luz_exp,agmon_exp} and  
followed by many attempts of its numerical estimation 
\cite{DFT_zundel,thecomputer,bond_order}, without a satisfactory 
accordance. 
Nevertheless, over the past years AIMD simulations have given many 
insights on the physics of Grotthuss mechanism 
\cite{thecomputer, parri_nature, pair_dance, chem_rev_pt,marco_prl, hassanali_PT}. 
It has been found that proton transport is
abnormally high and it is not driven by any ordinary diffusion
process.  
The picture which is instead commonly accepted describes the
translocation of excess positive charge along the  
hydrogen-bonded network as structural diffusion of the defects. 
Ref.~\onlinecite{hassanali_PT} has very recently shed new lights on 
proton transfer and has highlighted the complex 
nature of the process which involves both structural and 
dynamical changings of the hydrogen-bonded network.   

One of the main issues which sets back from a complete understanding of PT in
water, is related to the very sensitive 
thermal behavior of PT. The required PES precision of the
order of few
tenths of Kcal/mol has been reached only recently by the
state-of-the-art computational methods beyond DFT, such as coupled
cluster (CC), multi-reference configuration interaction (MRCI) and the
full configuration interaction quantum Monte Carlo (FCIQMC) methods;
on the other hand, they are characterized by a much poorer scalability
with respect to DFT and therefore they do not allow simulation of 
sufficiently large molecular clusters.

Over the past years, two main structural models has been 
proposed to explain proton hydration: M. Eigen 
\cite{eigen_model} considered as core of PT the complex 
$H_9O_4^+$ in which an hydronium $H_3O^+$ is strongly bounded
to three water molecules;
G. Zundel \cite{zundel_model} proposed a different reaction which
involves the simpler $H_5O_2^+$  complex (the ``Zundel ion'' or
``protonated water dimer'') as core
of the transfer of proton between two \htwoo\
molecules.

It has been demonstrated \cite{parri_nature} that these two models
occur only as limiting or ideal structures for a more complex
phenomenon which involves a large portion 
of the hydrogen-bonded network. 
Nevertheless they constitute a perfect testing ground for the most
advanced simulation techniques. In particular, the protonated water
dimer $H_5O_2^+$ is the smallest  
system in which an 
excess proton is shared between water molecules; due to its
simplicity, many studies have been carried out to elucidate
its properties. 
Over the past years, several accurate theoretical works have
appeared on the Zundel ion to study its structure and
energetics\cite{DFT_zundel,brueckner,aaauer,wales,extensive_H5O2};  
the fast development of spectroscopical instruments allowed to probe
experimentally vibrational properties of ionic species and therefore
many studies on this side 
have been published \cite{zundel_vibrational} on $H_5O_2^+$.Also
molecular dynamics simulations which include quantum effects
via Feynman path integrals have been performed  
(see Ref.~\onlinecite{parri_science} as a relevant example)
and recently an accurate PES has been produced \cite{huang} with
state-of-the-art coupled cluster technique including single, double and
perturbative triple excitations (CCSD(T)).

In the present paper, we report on an extensive study of the protonated
water dimer using a quantum Monte Carlo (QMC) approach based on a
highly correlated variational 
wave function for energy and geometry calculations. 
This approach has a very good scalability with the number of particles
with respect to 
other post-DFT methods. We demonstrate that it also
insures the precision of a few tenths of Kcal/mol required for an
accurate study of the system. These appealing targets are achieved by 
our Jastrow correlated antisymmetrized geminal power wave function (JAGP),
developed on an efficient contracted single particle basis set. While
the JAGP provides the necessary ingredients to treat both dynamical
and static correlations, the contracted basis set, based on atomic hybrid
orbitals, reduces dramatically the computational cost to
optimize the JAGP wave function.
Thanks to these promising features, 
we hope that it will pave the way to a complete understanding of the
underlined physics of PT and other properties of liquid water.

The paper is organized as follows.
In Sec~\ref{comp_methods} we introduce the computational techniques
used in this work. In particular, in Subsec.~\ref{QMC_wf} we describe
the variational ansatz and basis set which are the optimal compromise
between accuracy and computational cost for water clusters.
In Sec.~\ref{qmc_water_sec} we present our results on the single
\htwoo\ molecule, the subject of a vast amount of literature.
It represents thus the most suitable system  for a
first benchmark of our approach. For the single \htwoo,
we study the basis set and wave function convergence and show that we
have an accuracy comparable to the latest QMC studies.  
Sec.~\ref{water_dimer} is devoted to the main results of our paper on
the protonated water dimer. We performed structural calculations which
aim at finding the ground state geometry of $H_5O_2^+$ at  
the JAGP level of theory. In order to work out a zero-temperature
potential energy landscape, we chose the distance between the two
oxygens as a natural reaction coordinate.  Along this path, we investigated how the
symmetric ``Zundel'' configuration - namely the one with the proton evenly
shared between the two oxygens - evolves to a symmetry-broken
geometry, where the excess proton is more bounded to one of the two molecules.
We believe that this crossover is critical to understand the physics
of PT in larger water clusters. Here we show how our approach
describes this relevant region of the $H_5O_2^+$ PES.
The evaluation of proton transfer static barriers at fixed
$\overline{OO}$ distances shows that our QMC approach has a global
accuracy close to the state-of-the-art CCSD(T)  
calculations performed by Huang et al. \cite{huang}, with a discrepancy
of 0.2 Kcal/mol obtained by diffusion Monte Carlo (DMC) on the
variational Monte Carlo (VMC) geometries.
In conclusion (Sec.~\ref{conclusions}), we show that the favorable scalability
of our method allows one to simulate larger protonated water clusters
in a reasonable computational cost without loosing the accuracy 
reached in the smaller Zundel ion.
For the sake of test, we performed a structural optimization on a more
realistic PT model, i.e. the cluster of 6 water molecules and one
excess proton, to study the behavior of the computational
cost with the system size in practice.

\section{Computational Methods} 
\label{comp_methods}

\subsection{Density Functional Theory calculations}
\label{dft_section}

In this Section we present how we performed DFT-based calculations on
the Zundel ion. They are carried out to obtain optimized starting
geometries and wave functions for further quantum Monte Carlo calculations. 

The DFT geometry has been obtained by a Car-Parrinello dynamics
performed in a plane waves basis set with the \textsc{Quantum
  ESPRESSO}\cite{espresso} suite of codes. The technical details of
this calculation are reported in Subsec.~\ref{car_parri_dft}.

Once the ground-state DFT geometry is found, the Kohn-Sham (KS) orbitals, used as
starting input of our QMC wave function, are determined by the DFT
code of the \textsc{TurboRVB} package \cite{sorella_rvb} which works in a
localized Gaussian basis set, as described in Subsec.~\ref{gaussian_dft}.

\subsubsection{Plane waves DFT} 
\label{car_parri_dft}

With the aim at finding the optimal DFT geometry for further QMC
structural relaxation (see Sec.~\ref{qmc_methods}),  we use the 
Perdew-Burke-Ernzerhof \cite{pbe} (PBE) version of the generalized 
gradient approximation for the exchange-correlation functional. It
has been proven to be quite reliable in describing properties of
liquid water\cite{parri_jctc}.  
Core electrons are taken into account using a norm conserving
Trouiller-Martins pseudopotential\cite{trouil_mart}. The
Kohn-Sham orbitals are expanded  
over plane waves up to a cutoff of $37.5$ H for the wave function and
$150$ H for the charge density in a periodic box of 30 Bohr radii ($a_0$) per side.

The zero-temperature geometry relaxation is performed with
damped dynamics within the Car-Parrinello (CP) approach \cite{car_parri}. 
This method allows a very small force convergence threshold
of $\sim 10^{-5}$ H$/a_0$ on each atomic component at the end
of the relaxation.
The potential energy surface of the dimer is very flat \cite{extensive_H5O2}
and it develops on energy differences of less than 1
Kcal/mol. Therefore such an accuracy on atomic forces is essential for
the minimization to provide a reliable DFT description of the 
system. 

\subsubsection{Gaussian DFT} 
\label{gaussian_dft}

We perform gaussian DFT calculations with both pseudopotentials 
and full electron-ion potentials. 
The single-particle Kohn-Sham (KS) orbitals $ \{ \phi_i (\mathbf{r}) \}$ 
are used later on as optimized starting guess for the QMC wave function, 
as discussed in Sec.~\ref{qmc_wave_function_ansatz}.

The one-body DFT wave function is expanded over a primitive basis of 
\emph{Gaussian type orbitals} (GTO's) centered on the atom $a$,
defined up to a normalization constant as:
\begin{equation}
  \psi_{a,(l,m, n)}^{\textrm{GTO}} (\mathbf{r}) \propto
  |\mathbf{r}-\mathbf{R}_a|^l e^{-\zeta_{l,n}|\mathbf{r}-\mathbf{R}_a|^2} Z_{l,m}(\Omega_{\mathbf{r}-\mathbf{R}_a}), 
  \label{gaussians} 
\end{equation} 
with $m \in [-l,l]$ and $n \in [1,n_l]$,
where $\{ l,m \} $ are angular momentum quantum numbers, $n_l$ identifies the number of
Gaussians for each angular momentum shell, $Z_{l,m}(\Omega)$ are the spherical 
harmonics, and $\mathbf{r}$ and $\mathbf{R}_a$ are the electron and
ion positions, respectively. Our DFT basis set is \emph{even tempered}, 
namely for each angular
momentum $l$ the exponents of the Gaussians 
are expressed as a power series: $ \zeta_{l,n} = \alpha_l \beta_l^{n-1}$ for $n = 1,\dots,n_l$.
This allows
a reduction of the number of parameters,
as for each shell $l$ the full $\zeta_{l,n}$ series is fixed just by three
values: $\alpha_l$, $\beta_l$, and $n_l$. 
Denoting the set of GTO quantum numbers as $\mu = \{l,m,n\}$, the DFT orbitals $\{ \phi_i (\mathbf{r}) \}$ 
expanded over the primitive basis set read:
\begin{equation}
 \phi_i (\mathbf{r}) = \sum_a^{N_{\textrm{atoms}}}
 \sum_{\mu}^{N_\textrm{basis}(a)} c^i_{a,\mu} \psi_{a,\mu}^{\textrm{GTO}} (\mathbf{r}),
 \label{dft_expansion_primitive}
\end{equation}
where $N_{\textrm{atoms}}$ identifies the number of atoms of the
system, $N_\textrm{basis}(a)$ is the size of the GTO set of the
atom $a$. 
Thus the global DFT wave function can be expressed as the determinant
of the matrix $ \{\hat{\phi}_i(\mathbf{x}_j)\}$:
\begin{equation}
 \Psi_{\textrm{DFT}}(\mathbf{x}_1,\dots,\mathbf{x}_N) = \det \hat{\phi}_i(\mathbf{x}_j),
 \label{global_dft_wf}
\end{equation}
with $\{ \mathbf{x}_i = (\mathbf{r}_i,\sigma_i) \}$ the spatial and spin 
coordinates of the $N$ electrons in the system, and 
$\hat{\phi_i}(\mathbf{x}_j)=\phi_i(\mathbf{r}_j) \hat{\sigma_i}$ are the spin orbitals.

The ab-initio Hamiltonian of a quantum system is characterized by the divergence of the 
Coulomb potential at the electron-electron and electron-ion coalescence points.  
When the potential energy becomes infinite, the wave function of the
system must have a cusp to cure the corresponding singularity of the
Hamiltonian\cite{kato}. 
These singularities represent a serious issue when
an approximated wave function ansatz is employed to solve the
Schr\"odinger equation of a quantum system.

Within a DFT framework, the divergence created by electron-electron coalescence is 
intrinsically solved by the mean-field description of the system,
which maps the interacting Hamiltonian onto a one with independent electrons.
Nevertheless, the electron-ion coalescence still represents a problem. 
One way to cure the latter divergence is to employ a smooth pseudopotential.
The pseudo-interaction is chosen such that it eliminates the need of electron-ion cusps in the
wave function, by leading to a faster basis set convergence.
 
In this work we employ an alternative way to solve this issue. We introduce an additional factor 
to the DFT wave function ansatz which automatically satisfies the
electron-ion cusp conditions and cancels out the corresponding
divergences of the Hamiltonian.

We rewrite Eq.~\ref{global_dft_wf} as:
\begin{equation}
 \tilde{\Psi}_{\textrm{DFT}}(\mathbf{x}_1,\dots,\mathbf{x}_N) = J_1(\mathbf{r}_1,\dots,\mathbf{r}_N)  \det \hat{\phi}_i(\mathbf{x}_j),
 \label{corrected_dft_wf}
\end{equation}
where the function $J_1$ is called one-body Jastrow factor, borrowed
from the QMC notation.
It is defined by means of a simple function $u(\mathbf{r})$ which contains only one 
variational parameter $b$:
\begin{equation}
u(|\mathbf{r}-\mathbf{R}|) = \frac{1 - e^{-b |\mathbf{r}-\mathbf{R}|}}{2b}. \label{jas_u}
\end{equation}
The one-body Jastrow factor $J_1$ then reads:
\begin{equation}
 J_1 = \exp \left( -\sum_i^N \sum_j^{N_\textrm{atoms}} (2Z_j)^{3/4} u( (2Z_j)^{1/4} |\mathbf{r}_i - \mathbf{R}_j| ) \right), 
 \label{jas_dft}
\end{equation}
\newline
where $\mathbf{R}_j$ are the atomic coordinates corresponding to the atomic number $Z_j$. 
In the function $u$, the multiplicative factor $(2Z_j)^{3/4}$ is set
by the electron-ion cusp conditions, while the scaling length $(2Z_j)^{1/4}$ 
deals with the large distance behavior set by the random phase approximation\cite{holzman}.

Eq.~\ref{corrected_dft_wf} has the typical functional form of the Jastrow-single determinant 
(JSD) variational ansatz for quantum Monte Carlo calculations. 
Indeed the one-body Jastrow of Eq.~\ref{jas_dft} has been introduced in 
Ref.~\onlinecite{michele_benzene} to provide a solution of the electron-ion cusp 
conditions within a QMC framework.
Differently from QMC, our $\tilde{\Psi}_{\textrm{DFT}}$
presents only the one-body Jastrow factor since in DFT
the many-body electron-electron interaction 
is solved within a mean-field description.

By letting the wave function to fulfill the electron-ion cusp conditions, we are able to perform
calculations considering the bare interaction for both oxygen and hydrogen atoms (full potential 
calculations). 
Furthermore, at variance with conventional 
DFT calculations based on gaussian localized orbitals, in our approach
all the overlap matrix elements involved are computed  
as numerical integrals over a uniform mesh. Such integrals, due to the presence of the Jastrow 
factor, converge very rapidly with the number of points in the mesh and thus 
can be evaluated with a reasonable computational effort.\cite{azadi}

With the aim at reducing the computational cost, calculations are carried out also
by replacing the 1s core electrons of oxygen with the
pseudopotential approximation, while the hydrogen atoms are treated
always with the full Coulomb potential.

The oxygen pseudopotential that we used in both DFT and QMC
calculations is the Burkatzki-Filippi-Dolg (BFD) pseudopotential,
introduced in Ref.~\onlinecite{filippi_pseudo}. 
It is built by means of an energy-consistent approach in order to
reproduce the valence all-electron excitation energies for a number of
different atomic configurations computed at a scalar relativistic
Hartree-Fock (HF) level.

Gaussian DFT calculations are performed with a local density approximation 
(LDA) for the exchange-correlation functional. 
The optimal basis sets which approach the DFT complete basis set limit are
$O(8s8p5d)\,H(5s3p)$ and $O(9s9p6d)\,H(5s5p)$ for the pseudopotential and
full potential cases, respectively.  
Expanding the wave function over these basis sets, the variational parameters  
($\alpha_l$, $\beta_l$, $n_l$ for each angular momentum $l$ and  
the one-body Jastrow parameter $b$) are optimized at the
DFT level of theory, by minimizing the total LDA-DFT energy.

\subsection{Variational Quantum Monte Carlo calculations} 
\label{QMC_wf}

Quantum Monte Carlo (QMC) refers to several numerical techniques for 
electronic structure calculations of quantum systems. The interest 
of the scientific community for these methods has remarkably grown 
over the last three decades since they were successfully
applied to highly correlated electronic systems, and from small- to
medium-size quantum chemistry systems.

The QMC techniques used in this work are wave function-based, 
hence they aim at finding the many-body
wave function in a representation as close as possible to the 
true ground state of the system. The analytic form of the wave function
has to be square-integrable and computable
in a finite amount of time, but it does not present any further restrictions. 
This property allows us to introduce correlations in a compact and 
efficient way.

Another appealing feature of QMC methods is  the scaling with the number 
of particles. Indipendently of the technique, QMC provides an intrinsic scaling of 
$\sim N^3$-$N^4$, with $N$ the number of electrons, whereas
state-of-the-art coupled cluster single and 
double (CCSD) is $\propto N^6$ and coupled cluster single, double 
and triple (CCSD(T)) methods are $\propto N^7$. Therefore QMC 
approaches allow the treatment of larger molecular clusters with 
respect to other highly-correlated methods.

Major efforts have been done to reduce the multiplicative scaling
prefactor of QMC techniques, i.e. the cost of a single-point
calculation. The research has followed two main paths.  
From one side more compact trial wave functions have been developed.  
One of the most promising functions is the Jastrow-Antisymmetrized
Geminal Power (JAGP) used in this work. A description of the
wave function ansatz is given in Subec.~\ref{qmc_wave_function_ansatz}.

A second way to decrease the scaling prefactor relies 
on improving the quantum Monte Carlo estimators. 
QMC total energy calculations are in general much more efficient than those
of other observables, due to the \emph{zero-variance} property 
of the energy estimator.
Generally speaking, estimators of other important observables, as 
the charge density, do not possess this
property. Some other bare estimators, as the one for the atomic
forces, could have infinite variance.
Thus they affect the overall computational efficiency of 
the QMC calculation. In particular, over the recent years a major improvement 
has been achieved in the development of an efficient estimator for
nuclear forces\cite{estim1,estim2,estim3}. 

In the present work, we use a version of this estimator based on the
space-warp coordinate transformation\cite{umrigar_warp}
and implemented with an exact infinitesimal differentiation
method\cite{sorella_warp}.
This allows a single sample calculation of  the ionic forces 
with a computational effort of the order of $N^3$;  
this scaling is comparable to the one for total energy.
Moreover, the infinite variance of the space-warped force estimator in
the proximity of the nodal surface of the wave function is solved as
explained in Refs.~\onlinecite{attaccalite_sorella,zen_water}. The
latter Reference provides also a nice review on the latest progress
in the QMC nuclear forces evaluation.  
Therefore, by means of this state-of-the-art scheme force estimator we are able to
perform efficient structural optimizations 
at the QMC level of theory also for not-so-light atoms as oxygen.

\subsubsection{Wave function ansatz} 
\label{qmc_wave_function_ansatz}

The typical Quantum Monte Carlo wave function is made of
a symmetric bosonic factor (\emph{Jastrow factor} applied to 
 an antisymmetric fermionic part (\emph{determinantal part}):
\begin{equation}
  \Psi(\mathbf{x_1},\dots,\mathbf{x_N}) =
  J(\mathbf{r_1},\dots,\mathbf{r_N}) \times
  \psi_D(\mathbf{x_1},\dots,\mathbf{x_N}),
\label{total_qmc_wf}
\end{equation}
where the set $\{ \mathbf{x_i} = (\mathbf{r}_i,\sigma_i) \}$ represents spatial and spin 
coordinates of the electrons, as in Eq.~\ref{global_dft_wf}.

The Jastrow factor is a function of the electron-electron separation. 
It has been proven to be a crucial ingredient in order to well reproduce correlations
of the true many-body wave function; it contributes mainly in the description of 
dynamical correlation effects arising from 
charge fluctuations.
Furthermore, it has been shown \cite{michele_benzene} that the
Jastrow factor is particularly suitable in the treatment 
of Van der Waals intermolecular forces. They play an important
role also in the physics of the protonated water dimer, therefore 
an efficient parametrization of the Jastrow 
is essential.

We split this factor in one-body, two-body and three/four-body terms ($J=J_1 J_2 J_3$).

The one-body factor accounts for electron-ion interactions and it has been already
introduced in Eqs.~\ref{jas_u},\ref{jas_dft}. The two-body term deals
with the electron-electron interactions. 
In complete analogy with the one-body Jastrow, it is parametrized by a simple function 
of the electron-electron separation:
\begin{equation}
 u(|\mathbf{r}_i-\mathbf{r}_j|) = \frac{1 - e^{-b \abs{\mathbf{r}_i-\mathbf{r}_j}}}{2b}.
\end{equation}
Then it reads:
\begin{equation}
 J_2(\mathbf{r}_1 , \dots , \mathbf{r}_N) = \exp \left(  \sum_{i<j}^N u(\abs{\mathbf{r}_i-\mathbf{r}_j})   \right).
\end{equation}
$J_1$ and $J_2$ satisfy the Kato cusp conditions\cite{kato},
therefore correcting the divergence of the Coulomb potential energy at 
electron-ion  and electron-electron coalescence points, respectively.

The function $u$ rapidly decays to a constant value as the electron-ion and 
electron-electron distances increase; thus the large distance behavior 
of correlations is described by the three/four-body Jastrow term $J_3$.
It deals with electron-electron-ion (if $a=b$) and electron-ion-electron-ion 
(in the case $a\neq b$) correlations effects and it represents an essential 
part of our wave function; therefore we parametrize it in a richer way
than the other Jastrow terms:
      \begin{eqnarray} \label{3body}
      J_3(\textbf{r}_1,...,\textbf{r}_N) &=& \exp \left( \sum_{i<j}^{N} \Phi_J(\textbf{r}_i,\textbf{r}_j) \right) \nonumber \\
      \Phi_J(\textbf{r},\textbf{r}^{\prime}) &=& \sum_{a,b}^{N_{\textrm{atoms}}} \sum_{\mu,\nu}^{N_{\textrm{Jbasis}}}
      g_{\mu,\nu}^{a,b} \psi_{a,\mu}^J(\textbf{r})\psi_{b,\nu}^J(\textbf{r}^{\prime}),\nonumber
      \end{eqnarray}
where $N_{\textrm{Jbasis}}$ represents the number of GTO's of the primitive Jastrow basis for each atom.  
The Jastrow uncontracted orbitals  $\psi_{a,\mu}^J(\textbf{r})$ have
the same form as the primitive GTO basis set
$\psi^\textrm{GTO}_{a,\mu}(\textbf{r})$ for the determinantal part in
Eq.~\ref{gaussians}. The Jastrow primitive GTO basis
used in this work is $[O]3s2p1d[H]2s1p$.

The choice of the fermionic part of the wave function is more delicate.
In the case of spin unpolarized systems ($N_{\uparrow} = N_{\downarrow}$) of $N$ electrons, 
we can express it in a general way as an antisymmetrized product of the
\emph{geminals} or pairing functions $  \Phi(\mathbf{x}_i,\mathbf{x}_j) $ of the system:
\begin{eqnarray}
 \Psi_D (\mathbf{x}_1 \dots \mathbf{x}_N) & = & \hat{A}\left[ \Phi(\mathbf{x}_1,\mathbf{x}_2),\dots,\Phi(\mathbf{x}_{N-1},\mathbf{x}_N) \right] \nonumber \\
								& = & {\rm pf}(\Phi(\mathbf{x}_i,\mathbf{x}_j)).
\label{global_wf} 								
\end{eqnarray}
The geminals are antisymmetric functions of two electron
coordinates expressed
as a product of a spatial symmetric part and a spin singlet:
\begin{equation}
  \Phi(\mathbf{x}_i,\mathbf{x}_j) = \phi(\mathbf{r}_i,\mathbf{r}_j) \frac{\delta(\sigma_i,\uparrow)\delta(\sigma_j,\downarrow) - \delta(\sigma_i,\downarrow)\delta(\sigma_j,\uparrow)}{\sqrt{2}}.
  \nonumber
\end{equation}
The simplest, but compuationally most expensive expansion of the geminal is
over the \emph{uncontracted atomic orbitals}, and it reads:
\begin{equation}
 \phi( \mathbf{r}_i,\mathbf{r}_j ) = \sum_{a,b}^{N_{\textrm{atoms}}}
 \sum_{\mu,\nu}^{N_\textrm{basis}} \lambda_{\mu,\nu}^{a,b}
 \psi_{a,\mu}^{\textrm{GTO}}(\mathbf{r}_i) \psi_{ b, \nu }^{\textrm{GTO}}
 (\mathbf{r}_j), \label{atomic_expansion} 
\end{equation}
where the orbitals $\psi_{a,\mu}^{\textrm{GTO}}(\textbf{r})$ have the form reported in 
Eq.~\ref{gaussians} with $\mu = (l,m,n)$. 
$N_\textrm{basis}$ represents the number of primitive GTO's per atom.
The introduction of the many-body Jastrow factor $J$ in the total QMC
wave function (Eq.~\ref{total_qmc_wf}) allows a reduction of the
primitive basis set size without loss of accuracy. 
This reduction has several advantages in the QMC framework. 
At first it obviosly decreases the computational effort for the wave 
function calculation. Furthermore it also guarantees a more robust
energy minimization. 
Indeed, given the fact that QMC energy derivatives are noisy, using 
a more compact primitive basis set reduces its redundancy and helps 
in finding more quickly the global minimum since the number of the 
effective directions in the Hilbert space is smaller. 
The size-reduced primitive basis employed for the 
determinantal part is $O(5s5p2d)\,H(4s2p)$ and $O(6s6p2d)\,H(3s2p)$
for pseudopotential and all-electron calculations, respectively. 
The number of variational parameters expanded over GTO's 
atomic orbitals is $P^\textrm{AGP} \propto N_\textrm{total basis}
\times (N_\textrm{total basis}+1)$, where $N_\textrm{total basis}$ is
the total number of GTO's in the whole system.

In this work we use two distinct functional forms for $\psi_D$. They are 
distinguished by the number of non zero eigenvalues - i.e. the rank -
of the matrix $\{ \lambda_{\mu,\nu}^{a,b} \}$ 
in Eq.~\ref{atomic_expansion}.

The first one is commonly referred as single determinant (SD) and it is 
recovered when the $\{ \lambda_{\mu,\nu}^{a,b} \}$ matrix has the
lowest possible rank compatible with the number of electrons in the
system, namely when its rank is equal to $N/2$. 
It can be shown that it is equivalent to the Slater determinant ansatz for 
HF calculations. In this case the general expansion (\ref{atomic_expansion}),
after diagonalization of the matrix $\{ \lambda_{\mu,\nu}^{a,b} \}$, 
is written as:
\begin{equation}
    \phi( \mathbf{r}_i,\mathbf{r}_j ) = \sum_k^{N/2} \lambda_k 
    \psi_k^{MO}(\mathbf{r}_i) \psi_k^{MO}(\mathbf{r}_j), \label{mol_orb}
\end{equation}
where $\psi_k^{MO}(\mathbf{r}) = \sum_{a,\mu} c^k_{a,\mu}\psi_{a,\mu}^{\textrm{GTO}}(\mathbf{r}) $ 
are \emph{molecular orbitals} (MO). In this work we obtain them starting
from the optimized Kohn-Sham orbitals of Gaussian DFT calculations discussed
in Sec.~\ref{gaussian_dft}. The number of variational parameters in this case is 
$P^\textrm{SD} \propto N/2 \times N_\textrm{total basis}$ where
$N_\textrm{total basis}$ is the number of linear coefficients for each
MO, equal to the total number of GTO's.
Close to the CBS limit $N_\textrm{total basis} >> N$, and so in the SD
there is a significant reduction of the number of parameters with respect to the
fully uncontracted AGP expansion in Eq.~\ref{global_wf}.  
Multiplying the SD by the Jastrow factor one obtains
the JSD wave function, which is optimized simultaneously in both the J
and SD parts.

The SD represents a particular limit of the function in Eq.~\ref{global_wf}. 
Letting the rank of $\{ \lambda_{\mu,\nu}^{a,b} \}$ be greater than $N/2$, 
one introduces multiconfigurational states and goes beyond the single determinant 
representation. In this more general case the determinantal wave function 
is called Antisymmetrized Geminal Power (AGP).

The AGP is the particle-conserving version of the Bardeen-Cooper-Schrieffer (BCS)
wave function and it accounts for static correlations in the system, namely deriving from 
nearly-degenerate electronic energy levels. Together with the Jastrow factor it forms the 
Jastrow-Antisymmetrized Geminal Power (JAGP) wave function which represents a 
practical implementation of the resonance valence bond idea introduced by Linus
Pauling for chemical systems \cite{pauli}. 
The JAGP wave function has been proven to be particulary accurate in describing
a wide range of strongly correlated systems 
\cite{michele_agp1,michele_agp2,sorella_hb,michele_benzene},
and it is the second variational form we tested in our work after the JSD.

In order to benefit from the AGP ansatz without paying the cost of
dealing with a too large number of variational parameters, which will
make the calculation unfeasible for big systems, we 
develop the AGP expansion on a contracted basis set. We used atomic
hybrid orbitals, also employed in Ref.~\onlinecite{zen_water}, as
contractions of the primitive GTO's. 
One of the features of the AGP ansatz (without Jastrow) is that the
geminal is also the one-body density matrix of the
system. With the aim at finding the most effective local (atomic)
basis set to describe the whole system, we project the full one-body
density matrix on its local atomic constituents, by retaining in the
expansion of Eq.~\ref{atomic_expansion} only the
$\lambda_{\mu,\nu}^{a,b}$'s with $a=b$, and setting to zero the other
terms. 
By diagonalizing the projected one-body density matrix, we obtain a set
of local natural orbitals for each atomic fragment. The atomic
orbitals obtained in this way are hybrid (i.e. linear combination of
primitive Gaussians not restricted to a given angular momentum shell),
as they describe the hybridization arising from the atomic embedding,
which breaks the spherical symmetry around the nucleus. We call them
\emph{atomic hybrid orbitals}.
Thanks to this feature, for each atom we keep not only the information on
the local electronic structure due to the nuclear charge, but also the
information of the nuclear embedding in the compound, namely the
impact of the environment on its electronic structure. 

The initial AGP useful to fix the hybrid basis can be determined by
DFT calculations (in that case the rank of the AGP is N/2 as it comes
directly from an SD wave function), or by a previously optimized
JAGP wave funtion (in the latter case the Jastrow factor is disregarded and
only the determinantal part is taken in the one-body density matrix
determination). Once the hybrid basis set is chosen, and the AGP
expanded upon it, the geminal reads:
 \begin{equation}
       \phi( \mathbf{r}_i,\mathbf{r}_j ) =
       \sum_{a,b}^{N_\textrm{atoms}}
       \sum_{\alpha,\beta}^{N_\textrm{hyb}}
       \tilde{\lambda}_{\alpha,\beta}^{a,b}
       \psi_{a,\alpha}^{hyb}(\mathbf{r}_i)
       \psi_{b,\beta}^{hyb}(\mathbf{r}_j), \label{hyb_orb} 
 \end{equation}
where $\psi_{a, \alpha}^{\textrm{hyb}}(\mathbf{r}) = \sum_{\mu} c^\alpha_{\mu} \psi_{a,\mu}^{\textrm{GTO}}(r_{ia}) $ 
are the contracted atomic hybrid orbitals, and $\mu =(l,m,n)$ as
before. $N_\textrm{hyb}$ is the number of atomic hybrid orbitals
required for an accurate description of each atom. After their first
determination by DFT or by previous JAGP calculations, they are
further optimized in the QMC energy minimization. 

The number of 
variational parameters of the wave function is
$P^\textrm{hyb} \propto N_\textrm{total hyb}^2 + N_\textrm{basis}
\times N_\textrm{total hyb}$, where $N_\textrm{total hyb}$ is the
total nuber of hybrid Gaussians in the whole system.
Since $N \approx N_{\textrm{total hyb}} <<
N_{\textrm{total basis}}$, the hybrid orbitals represent
the optimal basis set, which reduces at most the total number of
parameters in the correlated AGP framework. In order to find the best
value of $N_\textrm{hyb}$ for oxygen and hydrogen, we carried out a detailed
analysis of the variational energy versus the hybrid basis size for
the single water molecule,
reported in Sec.~\ref{qmc_water_sec}.

\subsubsection{Quantum Monte Carlo methods} 
\label{qmc_methods}

All QMC calculations of this work have been carried out with the 
\textsc{TurboRVB} program\cite{sorella_rvb}. 
 
All the variational parameters of the JSD and JAGP wave functions have
been optimized by means of the stochastic reconfiguration method with
Hessian accelerator, also called ``linear
method''\cite{SR1,SR2,SR3,SR4,michele_benzene}. 
The forces on the parameters and ionic positions have always 
finite variance thanks to a reweighting scheme for finite systems
introduced in Ref.~\onlinecite{zen_water} to cure the variance
explosion around the nodes of the wave function.

A systematic way of improving the quality of the VMC ansatz 
is to perform lattice regularized diffusion Monte Carlo (LRDMC)
calculations\cite{michele_lrdmc,filippi_lrdmc}, a projective technique which filters out the high-energy 
components yielding a more accurate evaluation of the correlation energy. In
the case of non-local pseudopotentials, as the one for the oxygen, LRDMC
goes beyond the locality approximation, by providing always
a variational upper bound of the true ground state energy.

In the present paper, we report on both VMC and LRDMC calculations
of the protonated water dimer.
The BFD energy-consistent pseudopotential described in Sec.~\ref{gaussian_dft} has been
employed to replace the oxygen-core 1s electrons. Statistical error bars are kept 
smaller than $0.1$ Kcal/mol in all QMC final calculation.

\section{Results and discussion}
\label{results}
We report the results of our QMC study of protonated water dimer. 
This Section is organized as follows. 

The first part (Sec.~\ref{qmc_water_sec}) is devoted to
benchmark computations on water molecule which aim at proving
the quality of our QMC ansatz. 

The second part (Sec.~\ref{water_dimer}) presents the QMC study on the
protonated water dimer.
Different level of theory are compared with QMC calculations
for both geometry and energetics outcomes in order to assess
the accuracy of our approach in the study of proton transfer
systems.

\subsection{Benchmark calculations of the $H_2O$ molecule} 
\label{qmc_water_sec}

The single water molcule has been the subject of many numerical
studies based on several quantum chemistry methods. It is essential
to have a good description of its structural and electronic 
properties in order to tackle the study of larger water clusters, as
the intramolecular degrees of freedom will significantly affect the
intermolecular environment due to the large water dipole moment and the
strong directionality of the H bond.

Here we report our pseudopotential and all-electron calculations on
the water molecule with different wave
function types and basis sets, with the aim at choosing the best
compromise bewteen accuracy and efficiency in order to transfer the
most convenient ansatz to the zundel complex and eventually to larger water
clusters.

\begin{table*}[!ht]
\begin{ruledtabular}
\caption {Pseudopotential calculations. VMC energies and number of parameters for the QMC wave functions
  used in this work of the single water molecule, fixed here at the
  experimental geometry. The BFD pseudopotential has been used for
  O. The primitive Gaussian basis 
  set is O(5s,5p,2d) and H(4s,2p) for the determinant. The number of
  total parameters varies depending on the type of contractions used in the
  determinantal part. The Jastrow functional form has been kept fixed
  and developed on a primitive Gaussian basis set of O(3s,2p,1d) and
  H(2s,1p). This gives a number of 213 Jastrow parameters, divided
  into 184 $g_{\mu,\nu}^{a,b}$, 9 $\zeta$ Gaussian exponents in the
  uncontracted basis, 1 two-body homogeous Jastrow factor
  coefficient and 1 parameter for the analogous one-body part. The
  other parameters come from the determinant, and are reported in the
  last set of columns.} \label{table_pseudo_water_energy}
\begin{tabular}{|r|d|d|d|r|r|r|}
  \multicolumn{1}{|r|}{Wave function ansatz}   & 
\multicolumn{3}{c|}{VMC energies} &
\multicolumn{3}{c|}{number of parameters} \\ 
\multicolumn{1}{|c|}{} & 
   \multicolumn{1}{c|}{Energy $E_x$ (H)} &
   \multicolumn{1}{c|}{Variance (H$^2$)}   &  \multicolumn{1}{c|}{$E_x
     - E_\textrm{JSD}$ (mH)} &
   \multicolumn{1}{c|}{$\lambda_{\alpha,\beta}^{a,b}$} &
   \multicolumn{1}{c|}{det orbitals} & \multicolumn{1}{c|}{total} \\
\hline
JSD    (uncontracted orbitals)   &             -17.24821(7)      &
0.2655(6) &  0.0 & 682 & 18 & 895\textsuperscript{\emph{a}}\\ 
JAGP (hybrid orbitals: 4O 1H)  &   -17.25013(8)      & 0.2635(12)  &
-1.91(11)  & 21 & 158 & 374\\ 
JAGP (hybrid orbitals: 4O 5H)  &   -17.25183(6)      &  0.2510(6) &
-3.62(10) & 105 & 238 & 538 \\
JAGP (hybrid orbitals: 8O 2H)  &   -17.25267(7)     &   0.2426(18) &
-4.46(10) & 78 & 298 & 571 \\
JAGP (hybrid orbitals: 8O 5H)  &   -17.25302(6)     &   0.2412(34) &
-4.89(10) & 171 & 358 & 724 \\
JAGP  (uncontracted orbitals)    &  -17.25389(6)   &  0.2296(5) &
-5.68(10) & 682 & 18 & 895 \\
\end{tabular}
\footnotetext[1]{Here the number of parameters is the same as the one
  in the JAGP wave function since in the JSD ansatz
  we rewrite the corresponding geminal (of rank $N/2$) on
  the uncontracted basis in order to optimize the MO's, as explained in
  Ref.~\onlinecite{michele_agp2}.}
\end{ruledtabular}
\end{table*}

\begin{table*}[!htp]
\begin{ruledtabular}
\caption {As Tab.~\ref{table_pseudo_water_energy}, but for all-electron
  calculations. We report here the VMC energies and number of
  parameters for the all-electron QMC wave functions of the single water molecule, taken at the
  QMC relaxed geometry. The geometries are reported in Tab.~\ref{table_geometry_H2O}.
 The primitive Gaussian basis set is O(6s,6p,2d) and H(4s,2p) for the determinant.
The number of total parameters varies depending on the type of contractions used in the
  determinantal part. The Jastrow functional form has been kept fixed
  and developed on a primitive Gaussian basis set of O(3s,3p,1d) and
  H(2s,1p). This gives a number of 418 Jastrow parameters, divided
  into 406 $g_{\mu,\nu}^{a,b}$, 10 $\zeta$ Gaussian exponents in the
  uncontracted basis, 1 two-body homogeous Jastrow factor
  coefficient and 1 parameter for the analogous one-body part. The
  other parameters come from the determinant, and are reported in the
  last set of columns.} \label{table_allel_water_energy}
\begin{tabular}{|r|d|d|d|r|r|r|}
  \multicolumn{1}{|r|}{Wave function ansatz}   & 
\multicolumn{3}{c|}{VMC energies} &
\multicolumn{3}{c|}{number of parameters} \\ 
\multicolumn{1}{|c|}{} & 
   \multicolumn{1}{c|}{Energy $E_x$ (H)} &
   \multicolumn{1}{c|}{Variance (H$^2$)}   &  \multicolumn{1}{c|}{$E_x
     - E_\textrm{JSD}$ (mH)} &
   \multicolumn{1}{c|}{$\lambda_{\alpha,\beta}^{a,b}$} &
   \multicolumn{1}{c|}{det orbitals} & \multicolumn{1}{c|}{total} \\
\hline
JSD    (uncontracted orbitals)   &             -76.40025(8)      &
1.412(3) &  0.0 & 1383 & 19 & 1819\\ 
JAGP (hybrid orbitals: 9O 2H)  &   -76.40504(9)    &   1.399(6)  &
-4.79(12) & 91 & 361 & 870 \\
JAGP  (uncontracted orbitals)    &  -76.40660(7)   &  1.374(3) &
-6.35(11) & 1383 & 19 & 1819 \\
\end{tabular}
\end{ruledtabular}
\end{table*}

\begin{table}[!htp]
\begin{ruledtabular}
\caption {Geometrical properties of the global minimum of the water
  molecule. We report a comparison between different QMC wave functions and
  experimental results\cite{exp_H2O_geometry}.} \label{table_geometry_H2O} 
\begin{tabular}{|r|d|d|}
 \multicolumn{1}{|r|}{}   & 
\multicolumn{1}{c|}{$\overline{OH}$ (\AA)} &
\multicolumn{1}{c|}{$\angle HOH$ ($^{\circ}$)} \\
\hline
pseudo JSD   & 0.9542(4) & 104.730 \\
pseudo JAGP & 0.9549(4) & 104.549  \\ 
all-electron JSD & 0.9539(4) & 105.187  \\
all-electron JAGP  & 0.9557(4) & 105.101 \\
experiment\cite{exp_H2O_geometry} & 0.95721(30)  & 104.522(50) \\
\end{tabular}
\end{ruledtabular}
\end{table}

\begin{table}[!htp]
\begin{ruledtabular}
\caption {LRDMC energy results extrapolated to the zero lattice space
  limit. The LRDMC calculations are performed in the fixed-node
  approximation. In the last row, we compute the energy gain due to a better nodal description
  provided by the JAGP wave function with respect to the JSD one. Note
that the agreement between the pseudopotential and all-electron
calculations has an accuracy of the order of 0.1 mH, despite their very different
total energies.} \label{table_lrdmc_water_energy}
\begin{tabular}{|r|d|d|}
  \multicolumn{1}{|r|}{}   & 
\multicolumn{1}{c|}{pseudo} &
\multicolumn{1}{c|}{all-electron} 
\\ 
\hline
$E_\textrm{JSD}$ (H)   & -17.26280(6) & -76.42475(15) \\
$E_\textrm{JAGP}$ (H) & -17.26528(6) & -76.42690(14) \\
$E_\textrm{JAGP} - E_\textrm{JSD}$ (mH) &  -2.48(9) &  -2.15(21) \\
\end{tabular}
\end{ruledtabular}
\end{table}

\subsubsection{Pseudopotential calculations}

As reported in Sec.~\ref{comp_methods}, the BFD pseudopotential\cite{filippi_pseudo} has
been used for oxygen, while the two hydrogens have been treated
all-electron. The primitive Gaussian basis set is O(5s,5p,2d) and H(4s,2p) for
the determinant, while for the Jastrow factor it is O(3s,2p,1d) and
H(2s,1p). Note that the latter basis set has been recently claimed to
be one of the most accurate in an extensive 
QMC study of single molecule water properties\cite{zen_water}, which used the same Jastrow
ansatz as ours. 
The inhomogeneous three- and four-body Jastrow term (Eq.~\ref{3body}) has
been developed directly on the uncontracted primitive basis set, whose flexibility
guarantees a very good description of dynamical correlations. 
On the other hand, for the antisymmetric part we tested two main wave
function forms, the single Slater determinant (obtained by using a
geminal with rank $N/2$), and the AGP function. The difference in
energy between the JSD and the JAGP wave functions, reported in
Tab.~\ref{table_pseudo_water_energy}, shows the size of static
correlations in the system, which amounts to 5-6 mH. This leads
also to better geometrical properties, as seen in
Tab.~\ref{table_geometry_H2O}. The JAGP geometry is closer to the
experiment than the JSD one, in both the OH distance and the HOH
angle. The structural effects of static correlations in the water
molecule have been already pointed out in Ref.~\onlinecite{zen_water},
where they were attributed mainly to a change in the local description
of the oxygen atom. In order to analyze in deeper detail the role of the 
AGP correlations in the water molecule, we studied the natural orbital
occupations coming from the diagonalization of the geminal. One of the
appealing features of the AGP theory is that the geminal 
wave function is directly related to the one-body density matrix of
the system (without Jastrow factor). 
Its diagonalization yields
the molecular natural orbitals as eigenvectors and their weights
are related to the modulus of the AGP eigenvalues 
which are plotted in
Fig.~\ref{AGP_eigenvalues}. The Figure shows that indeed the orbitals
above the HOMO (highest occupied molecular orbital in the single
Slater determinant representation) have a sizable weight, with a
distribution which falls abruptly to zero only after the 40-th
orbital (gray area in the plot). This reflects the multi determinant
character of the water molecule, taken into account by the AGP
ansatz. We thus believe that, although the entanglement of quantum
levels at the origin of static correlations can come from the oxygen
atom, its impact in water has a genuine molecular character. Last
but not least, the multi determinant AGP representation leads to a
better description of the nodal surface of the true ground state, with
a gain of about 2.5 mH in the fixed node LRDMC energy with respect to
the one obtained by using the JSD as trial wave function, as reported in
Tab.~\ref{table_lrdmc_water_energy}. 

We turn now the attention on how to reduce the AGP basis set in an effective way.
So far, both the JSD and JAGP wave functions have been developed on the
uncontracted primitive basis in order to exploit at most its
flexibility. Thus, the total number of variational parameters is
895 (see the last column of Tab.~\ref{table_pseudo_water_energy}),
quite large for a single molecule, particularly if one would like to tackle 
the study of larger water clusters by means of QMC techniques. 
The most important limitation of this approach is that the number of variational parameters 
corresponding to the matrix elements $\lambda_{\alpha,\beta}^{a,b}$
increases clearly as the {\rm square} of the atomic basis size; therefore this
should be reduced at minimum in order to make this approach feasible 
for a large number of molecules.

To this porpouse, as explained in Sec.~\ref{qmc_wave_function_ansatz}, we define a new smaller basis set
by contracting the O(5s,5p,2d)/H(4s,2p) Gaussian primitive basis via atomic 
natural hybrid orbitals. Each atom in the system is described by its
own set of hybrid orbitals. We study how the size of the contracted hybrid
basis set affects the quality of the geminal expansion. We compare it
with its rigorous lowest energy limit provided by the uncontracted
JAGP reference previously computed. 
The results are reported in
Tab.~\ref{table_pseudo_water_energy}. The smallest basis set which
includes the 1s for H and the 2s and 2p orbitals for O, thus taking
into account the 2s2p near degeneracy at the atomic O level, is the
4O 1H hybrid set (in self-explaining notations). It
gives the poorest energy and variance among the hybrid basis sets
considered, though being lower than the JSD ansatz. The best energy is
obtained with the largest hybrid basis tried here, namely the 8O 5H
set. It recovers a large fraction of static correlations and its energy is
less than 1mH above the uncontracted JAGP one. However, the
price to pay is that the parameter reduction is weak, the total number
of parameters being close to the one of the full JAGP expansion (see
last column of Tab.~\ref{table_pseudo_water_energy}). Indeed, while the
number of $\lambda_{\alpha,\beta}^{a,b}$ is still significantly lower
than the one for the uncontracted basis set, nevertheless the parameters in the contracted
orbitals grow too much. The best compromise between efficiency, i.e. total number of
variational parameters, and accuracy, i.e. variational energy, is provided by the
8O 2H basis, as it yields a significant gain in energy
with a small/moderate number of parameters. 
This advantage will be remarkable for large number of atoms, 
as the number of variational parameters corresponding to this 
atomic natural hybrid orbitals grows only \emph{linearly} with the 
number of atoms; on the other hand the number of parameters corresponding 
to $\lambda_{\alpha,\beta}^{a,b}$,
grows instead quadratically, but it remains still affordable since 
it is dramatically reduced by this approach (see Tab.~\ref{table_allel_water_energy}). 

Finally, we study how the AGP spectrum changes with the 
contracted hybrid basis sets. Fig.~\ref{AGP_eigenvalues} shows that, after a
complete wave function optimization, the natural orbital eigenvalues
magnitude of the
hybrid AGP covers the $10^{-2}-10^{-4}$ range of
the fully uncontracted AGP expansion, except for the shortest 4O 1H
basis, which clearly spans a too small Hilbert space. Moreover,
we checked that the JAGP expanded on the optimal 8O 2H
basis gives the same fixed node LRDMC energy as the full JAGP,
signalling that the nodal surface is properly described even by the
hybrid 8O 2H contraction.
Therefore we are going to
use it in the study of the protonated water dimer presented after in
the paper. As
reported in the method Section, the gain in efficiency of the hybrid
basis set is expected to
be larger and larger as the system size increases, as the quadratic
growth of the $\lambda_{\alpha,\beta}^{a,b}$ with the number of atoms
depends strongly on the atomic basis size.

\subsubsection{All-electron calculations}

\begin{figure}
\includegraphics[width=\columnwidth]{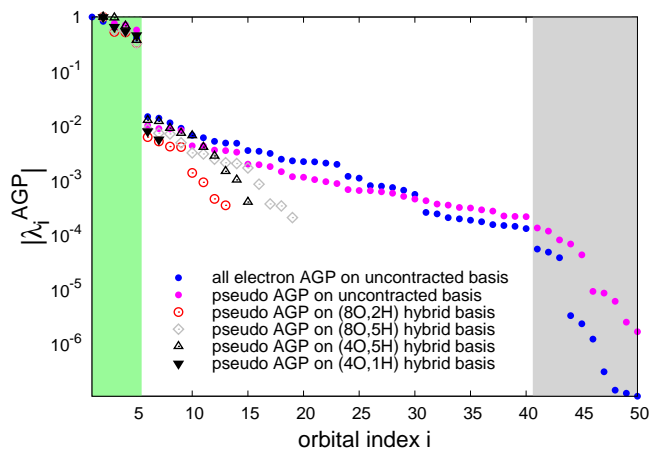}
\caption{Semilog plot of the modulus of the AGP eigenvalues
  versus the molecular orbital index for different basis sets
and calculations. The orbital indexes include always the oxygen 1s
electrons, replaced in the pseudopotential calculations.
The green area represents the exactly occupied molecular
orbitals in the single Slater determinant representation, with $\lambda_i=1$
for $i \in \{ 1, \ldots, \textrm{HOMO} \}$ and $\lambda_i=0$ for $i \ge \textrm{LUMO}$. In the
AGP representation, by diagonalizing the geminal we obtain the
corresponding molecular orbitals (eigenvecotors) and their
occupations $\lambda_i^\textrm{AGP}$ (eigenvalues). In the AGP,
also the orbitals above the HOMO are occupied, with a weight which
jumps across the HOMO to LUMO transition (in going from the green to
white region). The gray area shows when the MO occupation tail falls
rapidly to zero in the full AGP (expanded on a primitive basis set),
signaling that the MO's above that threashold start to be 
irrelevant to describe the static correlations in the system.} \label{AGP_eigenvalues}
\end{figure}

To study the accuracy of the pseudopotential approximation, we carried
out also all-electron calculations of the water molecule. With respect
to the pseudopotential calculations, the primitive and contracted basis sets for O have
been extended in order to account for the additional 1s electrons.
The primitive basis set is then O(6s6p2d) and O(3s3p1d) for the
determinant and Jastrow term respectively, while the optimal
contracted hybrid basis set is 9O 2H. 
In Tab.~\ref{table_allel_water_energy} we report the variational
energies for different wave functions. The energy gain provided by the
all-electron JAGP wave function is very close to the one in
pseudopotential calculations. The substantial agreement between the two
calculations is apparent also in
Fig.~\ref{AGP_eigenvalues}, where the eigenvalues of the
higher energy molecular natural orbitals in the AGP behave similarly. 
The LRDMC energy difference between the JAGP and JSD trial wave functions
coincides within the error bars with the one with pseudopotentials. The nodal contribution
to the fixed-node energy is the same. The JAGP LRDMC energy is one of
the best ever published in literature, in statistical agreement with
the one computed by Lüchow and Fink\cite{fink} (-76.429(1)) H), who
used 300 determinants in the trial wave function, and with the one by
Zen et al.\cite{zen_water} (-76.42660 H), who used the same ansatz as ours
in the trial. The JAGP LRDMC projected energy is only 11 mH higher than the
extrapolated exact result of −76.438 H\cite{exact_H2O_energy}.

The all-electron calculations confirm the importance of including static correlations to have a better
description of the geometry, as shown in
Tab.~\ref{table_geometry_H2O}. However the HOH angle turns out to be
less accurate than the one obtained with pseudopotentials if compared
to the experiment, most
probably because it is a quantity very sensitive to the basis set
convergence, which is harder to reach in all-electron calculations.
Another drawback of all-electron calculations is of course the larger variance
for an equivalent wave function ansatz, due to the 1s electron
fluctuations, as one can easily evince from the comparison between
Tab.~\ref{table_pseudo_water_energy} and Tab.~\ref{table_allel_water_energy}.

Due to the larger primitive basis set required in all-electron
calculations, the parameter reduction allowed by the hybrid basis contraction
(9O 2H) has a great impact on the efficiency. The total number of parameters is reduced by
almost a factor of 4 in the determinantal part of the single molecule,
without any significant loss of accuracy in the JAGP total energy (see
Tab. ~\ref{table_allel_water_energy}). 

The use of the BFD pseudopotential for oxygen, and the JAGP ansatz
together with the hybrid basis set, tested in the water molecule, is transferred to the
protonated water dimer.

\subsection{Protonated water dimer}
\label{water_dimer}
The protonated water dimer represents the simplest model 
for proton transfer in acqueous systems. 
In our work we focus on the energetics of $H_5O_2^+$ 
related to proton transfer by choosing suitable reaction 
coordinate (RC).
Selecting a RC allows to reduce the complexity of the 
full-dimensional PES. It projects 
it onto a single-dimensional subset which retains the 
most important physical features of the full hypersurface.

Since it is not unique, a correct choice of RC is essential 
in order to filter out the physical features which suitably 
describe the considered phenomenon. 
We propose a mechanism of proton transfer within the dimer
which leads to a natural definition of our RC.
 
As modeled by our system, the 
proton transfer reaction takes place in three different 
steps. At first the excess proton is bounded to one water molecule
forming an $H_3O^+ + H_2O$ complex. 
By thermal fluctuations
the oxygen-oxygen separation can get closer to the optimal distance 
of the Zundel complex (around $2.39$ \AA); at this stage the system assumes a
``Zundel configuration'' with  
the proton equally shared between the two oxygens. A further 
stretch of $\overline{OO}$ disfavors the Zundel
configuration and a new $H_2O + H_3O^+$ complex is produced. 
The overall effect of this process is a transfer of a proton
along the hydrogen bond between two oxygen atoms. This mechanism 
suggests to choose the oxygen-oxygen separation as RC
for the dimer potential energy curve.

We present in the following the geometry and energetics of the 
protonated water dimer. 
Results on the global minimum geometry are reported at first.
The potential energy curve is the result of a stretching of
the oxygen-oxygen distance; the
behavior of the excess proton is thus investigated 
along the same path.
A particular attention is devoted to
the behavior of the $H_3O^+ + H_2O$ complex in the
broken-symmetry region of the energy landscape.
Proton transfer static barriers at different $\overline{OO}$
distances are calculated in order 
to further estimate the accuracy of our approach. 
All QMC calculations are performed with the 
JAGP wave function ansatz developed on an atomic hybrid basis set, as
discussed in the previous Sec.~\ref{qmc_water_sec}.

\subsubsection{Properties of the symmetric global minimum}
\label{global_minimum}

\begin{figure*}
     \caption{QMC optimized geometries for global $C_2$ minimum (left) and for $C_s$ local minimum (right).} \label{geometries}
    \resizebox{1\textwidth}{!}{ \includegraphics[scale=1]{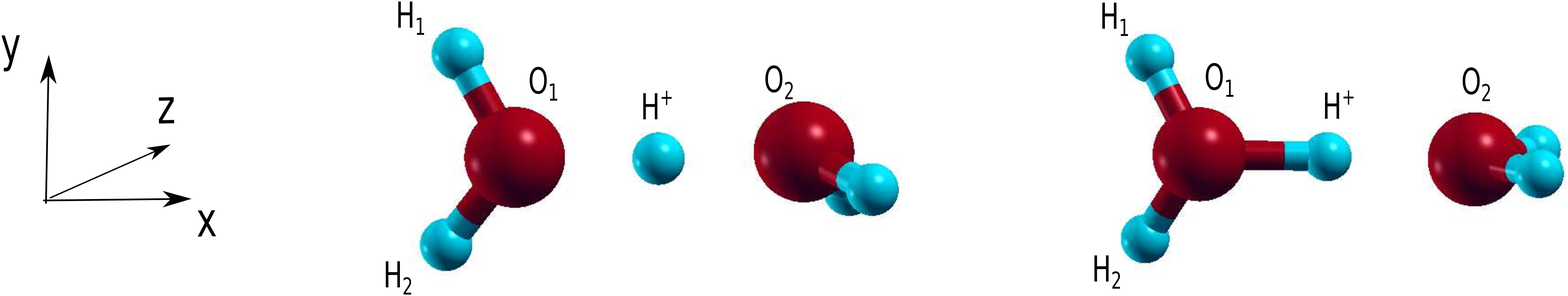}}
\end{figure*}
The minimum energy structure of $H_5O_2^+$ has been debated
in the scientific community, as there are two candidates with
competing energies: 
a $C_2$-symmetric structure, commonly addressed as Zundel 
configuration, with the proton evenly shared between 
the two oxygens, and a $C_s$-symmetric one with the proton slightly
localized on one $H_2O$ molecule (see Fig.~\ref{geometries}).
Several \emph{ab-initio} investigations 
\cite{brueckner, extensive_H5O2} have shown that a 
better treatment of electron correlations turns out in a 
change of the ground state (GS) geometry from the $C_s$-sym to the 
$C_2$-sym configuration. Accurate highly-correlated 
studies \cite{wales,aaauer,extensive_H5O2,huang} 
has eventually confirmed that the global minimum 
is $C_2$ symmetric.

At a QMC level of theory, the $H_5O_2^+$ GS 
shows a $C_2$ Zundel configuration with centrosymmetric excess 
proton (left part of Figure \ref{geometries}), in agreement with 
the previous studies.
The main geometrical parameters of the global minimum are 
presented in Tab.~\ref{table_geometry_C2}, both for the
pseudopotential and all-electron calculations.

\begin{table*}[!htp]
\caption {Geometrical properties (distances in \AA, angles in $^{\circ}$) of the $C_2$-symmetry minimum of protonated water dimer, comparison 
between different computational methods. See Fig.~\ref{geometries} (left-hand side) for atoms notation.} \label{table_geometry_C2}
\begin{ruledtabular}
\begin{tabular}{|c|d|d|d|d|d|d|}
  & \multicolumn{1}{c|}{$\overline{O_1O_2}$}   & \multicolumn{1}{c|}{$\overline{O_1H^+}$} & \multicolumn{1}{c|}{$\overline{H^+O_2}$}   &  \multicolumn{1}{c|}{$\angle O_1H^+O_2$} & \multicolumn{1}{c|}{$\overline{O_1H_1}$} & \multicolumn{1}{c|}{$\overline{O_1H_2}$} \\
\hline
DFT-PBE & 2.4111  & 1.2074   & 1.2074  & 173.661 & 0.9697 & 0.9691 \\

DFT - B3LYP\cite{wales} &   &  & 1.2172 & 173.6 & 0.9706 & 0.9701 \\ 

QMC - with pseudo & 2.3847(5) & 1.1930(5) & 1.1942(8)   & 174.71(7) & 0.9605(8) & 0.9650(8) \\

QMC - all-electron & 2.3905(4) &  1.1944(6) & 1.1989(5)  & 174.43(9)  & 0.9630(7) & 0.9628(6) \\

CCSD(T)\cite{huang} & 2.3864  & 1.1950  & 1.1950 &  173.730 & 0.9686 & 0.9682 \\
\end{tabular}
\end{ruledtabular}
\end{table*}

\begin{table*}[!htp]
\caption {Geometrical properties (distances in \AA, angles in $^{\circ}$) of the $C_s$-symmetry minimum of protonated water dimer. See Fig.~\ref{geometries} (right-hand side) for atoms notation.} \label{table_geometry_Cs}
\begin{ruledtabular}
\begin{tabular}{|c|d|d|d|d|d|d|}
  & \multicolumn{1}{c|}{$\overline{O_1O_2}$}   & \multicolumn{1}{c|}{$\overline{O_1H^+}$} & \multicolumn{1}{c|}{$\overline{H^+O_2}$}   &  \multicolumn{1}{c|}{$\angle O_1H^+O_2$} & \multicolumn{1}{c|}{$\overline{O_1H_1}$} & \multicolumn{1}{c|}{$\overline{O_1H_2}$} \\
\hline

DFT - B3LYP\cite{wales} &   &  & 1.2507 & 175.4 & 0.9746 & 0.9741 \\

QMC - with pseudo & 2.3996(6) & 1.1154(8)  & 1.2852(4)  & 176.5(1)  & 0.9641(7) & 0.9625(4) \\

QMC - all-electron & 2.3913(3) & 1.1285(5) &  1.2648(4) & 175.29(6)  & 0.9635(4) & 0.9616(5) \\

CCSD(T)\cite{huang} & 2.3989 & 1.1233 &1.2720 & 175.646 & 0.9641 & 0.9645 \\
\end{tabular}
\end{ruledtabular}
\end{table*}
If compared with CCSD(T), the QMC ground state geometries 
show an agreement of up to $0.005$ \AA\ in the atomic separations, and up to 
$1^{\circ}$ in the angles. 
As the angle is close to
$180^{\circ}$, in this case its determination is more delicate and could be
affected by a larger statistical bias. Therefore we notice a slight
discrepancy on this quantity between QMC and coupled cluster outcomes. 
Nevertheless, these differences do not affect 
the overall description of the GS and the energetics 
of the system.
In the present work, also the $C_s$-symmetric structure 
has been taken into account (left-hand side of Fig.~\ref{geometries}).
Tab.~\ref{table_geometry_Cs}, which reports the VMC optimized
$C_s$  geometries, confirms the trend seen in
Tab.~\ref{table_geometry_C2} for the GS, although the discrepances in
the bond lengths are slightly larger between different methods,
the CCSD(T) values being in between the all-electron and
pseudopotential VMC results. 

The energy difference between the $C_s$ configuration and
the $C_2$-symmetric global minimum turns out to be $0.25(8)$ Kcal/mol 
at VMC level and $0.23(8)$ Kcal/mol with LRDMC. 
This is in satisfactory agreement with previous 
results carried out 
with M\o ller-Plesset pertubation theory\cite{aaauer} ($0.28$ Kcal/mol) 
and CCSD(T)\cite{huang} ($0.46$ Kcal/mol) techniques.
The quality of pseudopotential is hence verified also for the Zundel
complex. In view of this, its PES along the chosen RC will be worked
out mainly with this approximation due to its substantially lower
computational demand.

\subsubsection{Stretching the $\overline{OO}$ distance}
\label{stretching}

\begin{figure}
     \caption{Potential energy curve (Kcal/mol) of the protonated water dimer projected on the $\overline{OO}$ distance. Comparison between 
     different computational methods. Structural relaxation is performed at each level of theory. Each curve has its minimum as reference point.} \label{landscape1}
    \includegraphics[width=\columnwidth]{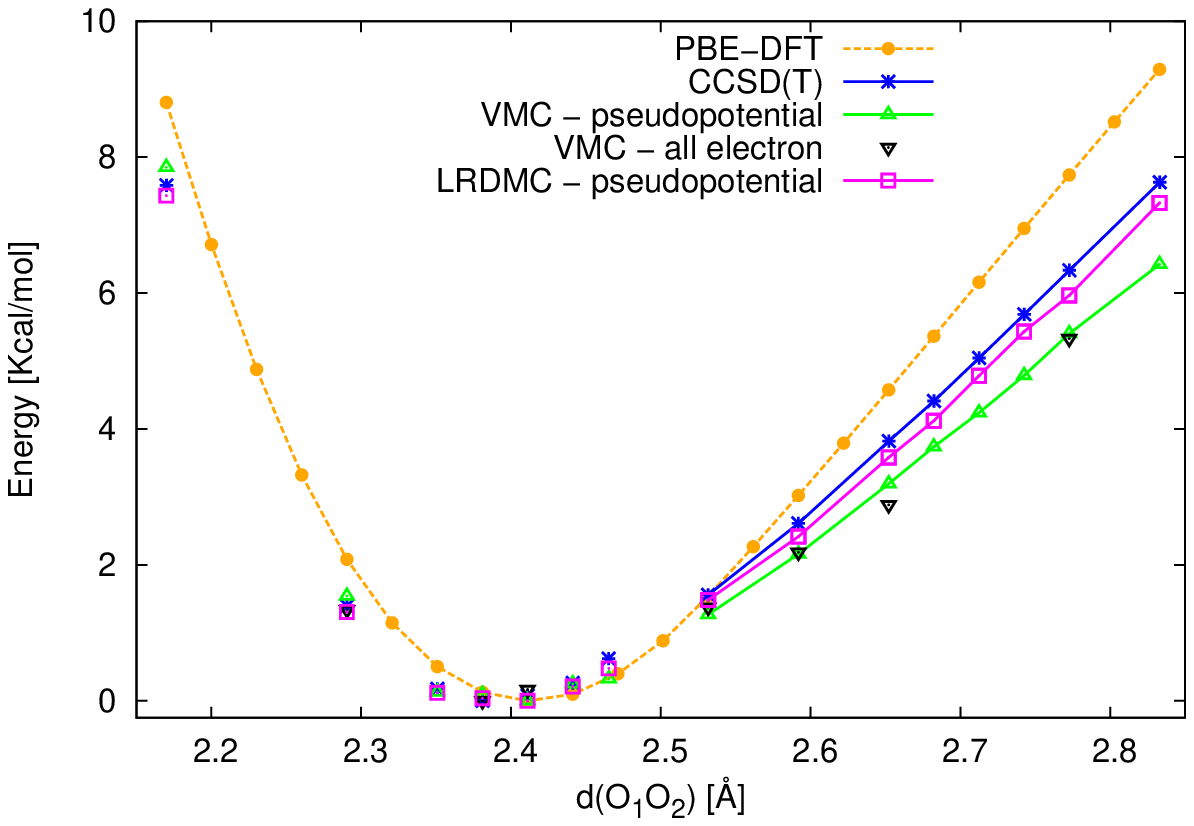}
\end{figure}

\begin{figure}
    \caption{CCSD(T) and VMC energy landscape (Kcal/mol) as a
      function of the $\overline{OO}$ distance (\AA) with respect to the
      LRDMC energies (the zero of the y-axis). Full potential VMC
      results are reported for few points along with dissociation 
      energies. The green area represents the error bar 
      achieved in typical LRDMC run, i.e. $\sim$ 0.1 Kcal/mol.
      } \label{blowup_landscape}
    \hspace*{-0.8cm}
    \resizebox{1.1\columnwidth}{!}{ \includegraphics [scale=1]{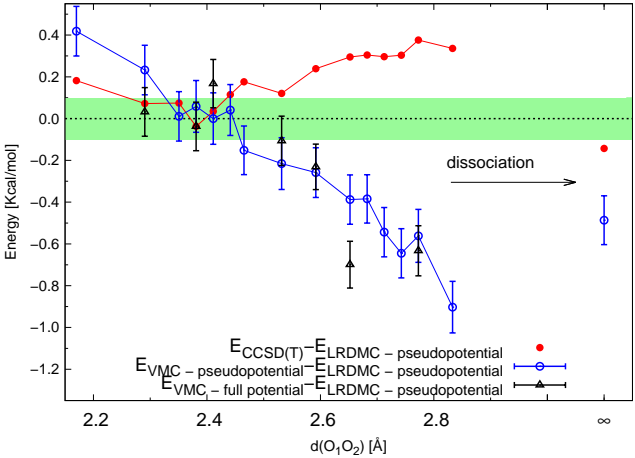}}
\end{figure}

From the global minimum $C_2$-sym configuration we stretch the oxygen-oxygen distance in order to 
study the potential energy curve and to elucidate the proton transfer properties in the dimer. 
The structural relaxation at the VMC level for fixed $\overline{OO}$ separation 
requires a careful procedure due to the flatness of the PES.
Starting from a PBE-DFT optimized geometry (see Sec.~\ref{car_parri_dft} for
technical details) and a JAGP variational wave function fully optimized in the
electronic part, the atomic coordinates are relaxed with the
steepest descent method, considering them as additional variational parameters 
of the QMC wavefunction.

For the sake of comparison, we minimize a parametrized full-dimensional
PES fitted from CCSD(T) calculations\cite{huang} to find the best coupled cluster
estimates of energy and geometry. By means
of the \emph{downhill simplex} minimization technique,
we find the configuration of lowest CCSD(T) energy at the
same constrained $\overline{OO}$  distance as the corresponding QMC
and DFT calculations.

In Fig.~\ref{landscape1} we plot the energy landscape along the RC
for PBE-DFT, CCSD(T), VMC and LRDMC, the latter computed at the VMC geometry.
For VMC technique, we report also full potential calculations for some
$\overline{OO}$ separations. 
We notice a substantial agreement 
among all techniques in the region at the left of the 
global minimum of the curve, except for a rigid shift by 0.02 \AA
between the PBE-DFT results and the others. The
PBE $\overline{OO}$ minimum is indeed located at 2.41 \AA, while the
minimum of the other methods turns out to be at $\sim$ 2.39 \AA.
Hence this part of the energy curve is only slightly influenced by a better
treatment of correlations. 

On the other hand, the region at the right of the minimum, at
intermediate $\overline{OO}$ distances ($\geq 2.55$ \AA), displays a
different behavior. The PBE-DFT overestimates the slope of the curve with respect to 
the most accurate techniques. LRDMC, which yields the best QMC
correlation energy, shows a remarkably good agreement with the
state-of-the-art CCSD(T) results.
In particular all CCSD(T) outcomes are in the range of $\approx 0.3$
Kcal/mol, three times the statistical error of the LRDMC calculations, as
shown in Fig.~\ref{blowup_landscape}.

\begin{figure}
    \caption{Protonated water dimer energy landscape (Kcal/mol) as a function of $\overline{OO}$ distance (\AA). All the calculations are 
    performed with VMC-optimized geometry. The zero energy reference point corresponds to the minimum  of each curve.} \label{landscape2}
    \includegraphics[width=\columnwidth]{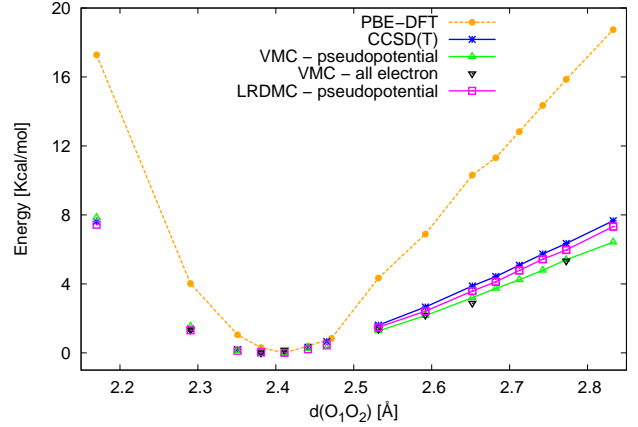}
\end{figure}

The curves reported in Fig.~\ref{landscape1} are obtained with the
minimum energy geometry at each level of theory (except for the LRDMC,
whose geometry is set at the VMC level).  
In order to obtain a more reliable comparison and avoid the bias
coming from the use of different geometries, we carried out the same
calculations employing the VMC-optimized structures for every
technique. The result is reported in Fig.~\ref{landscape2}. The trend
displayed in Fig.~\ref{landscape1} for method-optimized geometries is
remarkably enhanced when the same configuration of the dimer is
considered.  
Away from the minimum, the PBE-DFT energies show a larger
overestimation of the slope. 

\begin{figure}
    \caption{Separations (\AA) between the two oxygens and the excess proton as a function of the reaction coordinate for different computational methods.} \label{oh_distances}
    \includegraphics[width=\columnwidth]{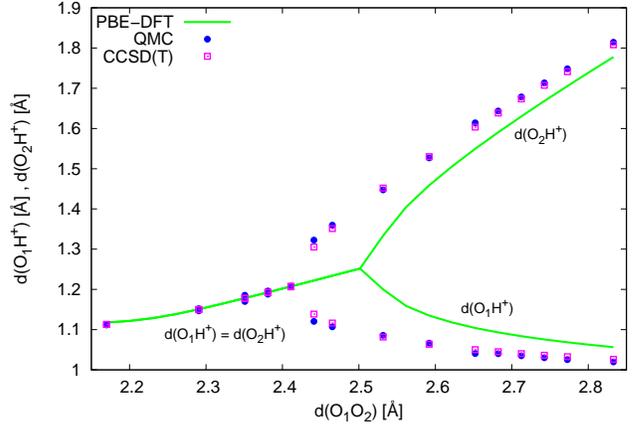}
\end{figure}

The slope of the dimer potential energy curve is related to the
behavior of the excess proton in the system. In
Fig.~\ref{oh_distances} this property is elucidated.  
We report in the same plot the separations between the excess proton
and each of the two oxygens: $\overline{O_1H^+}$ and $\overline{H^+O_2}$.  
The structure of the dimer is relaxed at different level of theories
with the same procedure carried out for the potential energy curve in
Fig.~\ref{landscape2}. The plot clearly shows  
the appeareance of two distinct regimes of the dimer. One is
characterized by a symmetric Zundel configuration with the proton evenly bonded
to the two oxygens; from the point of view of proton transfer physics,
it is basically equivalent to the GS configuration.
Stretching the $\overline{OO}$ distance results 
in the formation of a $H_3O^+ + H_2O$ complex with 
the proton localized on one water 
molecule. These configurations belong to the 
\emph{asymmetric regime} of the dimer.
Within this regime,
the initial $C_2$ point symmetry of the GS geometry
is broken due to proton localization.

Notice that the $C_s$-sym local minimum presented in Sec.~\ref{global_minimum}
displays a localized proton at a distance of $\sim 2.39$ \AA, smaller than
the symmetry-breaking distance shown in Fig.~\ref{oh_distances}. However,
we verified that if one stretches the $\overline{OO}$ distance 
starting from this local minimum, 
the energies obtained are higher than the values plotted 
in Fig.~\ref{landscape1} and Fig.~\ref{landscape2}.

The distances obtained by the QMC relaxation of the atomic coordinates
are in excellent agreement with the CCS(T) calculations; in particular
the root mean square distance between the two data sets
over the whole $\overline{OO}$ range is 
of $\sim 0.007$\,\AA\;for both $\overline{O_1H^+}$ and 
$\overline{O_2H^+}$. Electron correlation 
plays a key role in determining the stability of the symmetric
configuration of the dimer. The overestimation of the  
potential energy curve slope by PBE-DFT in Fig.~\ref{landscape1} 
corresponds to an overestimation of the symmetry-breaking
$\overline{OO}$ separation by $\sim 0.13$ \AA \, with respect to
higher-level post-DFT \emph{ab-initio} methods, and to a very poor description
od the geometry in the symmetry-broken region close to the
symmetry-breaking point, where the discrepancy in the
$\overline{OH^+}$ distances is the largest (going up to 0.15 \AA).

\subsubsection{Implications for more realistic PT models}

The zero-temperature potential energy curves reported in
Fig.~\ref{landscape1} and Fig.~\ref{landscape2}
seem to conflict with the proton transfer mechanism discussed in the
introduction of Sec.~\ref{water_dimer}, since the configuration with centrosymmetric
proton is energetically favored and therefore
it represents a stable state
rather than a transition state between two asymmetric 
configurations with localized proton.

This contraddiction is only apparent.
Indeed, Ref.~\onlinecite{parri_nature} has shown that the introduction
of thermal and polarization effects due to the physical environment,
favors the asymmetric regime of the complex. Indeed, 
at finite temperature the
\emph{free energy} landscape displays a global minimum
shifted towards the asymmetric regime
and the Zundel-like structure does not represent anymore 
the energetically favored configuration. 
Moreover, a recent experimental result shows that the average 
$\overline{OO}$ distance in liquid water is of 
$2.81$ \AA \cite{water_dist} which clearly corresponds to a 
symmetry-broken configuration of the dimer 
(see Fig.~\ref{oh_distances}).
Anyway, in order to jump
from a water molecule to one of its neighbors, the proton must pass through
a Zundel configuration. Describing correctly the energetics and
geometry of the protonated water dimer in the
symmetry-breaking transition region is therefore of paramount
importance to hope having an accurate description of the PT in more
realistic models.

Indeed, as already mentioned, Fig.~\ref{oh_distances} highlights that 
at $0$ K a better treatment of electronic correlations turns out
in a stability of the asymmetric regime over a considerably
wider range of
$\overline{OO}$ distances with respect to DFT results.  
Hence we expect that, at finite temperature and including the 
sorrounding physical environment, the value
of the free energy activation barrier for proton transfer 
would be significantly higher than the PBE-DFT prediction. 

\subsubsection{Properties of the symmetry-broken configurations}
\label{broken}

In this Section we focus the attention on some relevant properties 
of the broken-symmetry region where the excess proton is localized 
on a water molecule. At the QMC level of theory,
the formation of a $H_3O^+ + H_2O$ complex occurs at an oxygen-oxygen 
distance of $\sim 2.43$ \AA \, in perfect accordance with CCSD(T) results, 
as seen in Fig.~\ref{oh_distances}.

A quantity which has been 
extensively studied over the past years\cite{DFT_zundel,pavese,adamo, 
thecomputer} is the \emph{static proton transfer barrier}, i.e. the
barrier that the $H^+$ has to overcome in order to jump from one \htwoo\
molecule to the other at a fixed $\overline{OO}$ distance in the
symmetry-broken regime. This 
quantity does not provide a realistic comparison with the experimental 
activation barrier for PT, as the
$\overline{OO}$ distance will shorten during the proton hopping. 
Nevertheless, it is relevant in order to provide a further check of the accuracy 
of our QMC approach.
Fixing the $\overline{OO}$ distance, the barriers are 
obtained as difference between the
asymmetric configuration with localized proton and a structure with 
the excess proton in a centrosymmetric position. Calculations are performed
in three representative $\overline{OO}$ separations; the results are shown 
in Tab.~\ref{static_barrier} where they are compared with existent data 
in literature. 

The first $\overline{OO}$ distance is $2.47$ \AA, 
very close to the symmetry-breaking point of 
the dimer as displayed by highly correlated approaches; on the contrary, 
at a DFT level the configuration with centrosymmetric proton is still
energetically favored, as shown by Fig.\ref{oh_distances}.
It has been noticed\cite{extensive_H5O2} that the PES in this
region is particulary flat. 
Furthermore the potential energy curve of the dimer develops on very 
tiny energy differences around the symmetry-breaking point. 
These issues make the calculations in this region of the PES
extremely delicate since the stochastic noise can considerably affect 
the quality of the QMC predictions.

Indeed, Tab.~\ref{static_barrier} shows that QMC and CCSD(T) 
display a vanishing energy barrier of
the order of $0.1$ Kcal/mol as the hydrogen is displaced
along the oxygen-oxygen axis. 
The height of this barrier is slightly above the attained
statistical error in our typical QMC run. However, despite the very 
sensitive behavior of the dimer PES around the symmetry-breaking
point, the accuracy of our force minimization algorithm allows
to account for the tiny energy differences involved; thus it ensures
the necessary precision to describe the PT physics in the dimer.
As we discuss in the next Section, a similar accuracy can be
achieved in larger molecular clusters with a reasonable amount of 
computational time. Therefore our QMC framework guarantees 
a reliable description of the PT physics also for more realistic
models.

\begin{table}[!htp]
\caption {Static proton transfer barriers (Kcal/mol) at fixed
  $\overline{OO}$ separations. Comparison between different level of
  theory. }\label{static_barrier}
\begin{ruledtabular}
\begin{tabular}{|c|c|c|c|}
Method & \multicolumn{3}{c|}{$O-O$ distance} \\
\hline
  & $2.47$ \AA & $2.6$ \AA  & $2.7$ \AA  \\
\hline
DFT-PBE &  -0.74  & 0.19  & 1.94   \\

DFT-PBE (\emph{best}) \cite{adamo} & & 0.21 & 1.46 \\

VMC  & 0.28(6) & 2.99(8) & 5.99(8)   \\

LRDMC & 0.37(8) & 2.64(7) & 5.57(7)  \\

CCSD(T) - CCSD(T) geometry & 0.22 & 2.37 & 5.24 \\

CCSD(T) - QMC geometry  & 0.28  & 2.29 & 5.32  \\

CCSD(T) - MP2 geometry \cite{adamo} &  & 2.08 & 4.85 \\

QCISD(T) - MP2 geometry  \cite{adamo} & & 2.06 & 4.82  \\

MS-EVB \cite{thecomputer} & & 2.05 & 5.11 \\ 

MP2 \cite{adamo} &  & 1.77 & 4.39  \\
\end{tabular}
\end{ruledtabular}
\end{table}

The other results are obtained at larger oxygens separations, further from
the symmetry-breaking point. They confirm the general behavior already seen along the
$\overline{OO}$ RC.
LRDMC and CC results are in a good agreement up 
to $0.2-0.3$ Kcal/mol, whereas the VMC slightly overestimates 
the barrier. As well known from previous works, 
DFT substantially underestimates the barrier with respect to post-DFT
methods which provide a better treatment of correlations.

Finally, let us analyze the extreme limit of the asymmetric Zundel
configuration, namely when $\overline{OO} \rightarrow \infty$, with
the formation of one \htwoo\ and one hydronium.
The dissociation energy $D_e$ of $H_5O_2^+$ is computed by setting the distance
between the two oxygens to $14$ \AA. With the CCSD(T) PES, we checked that this is already
in the large distance plateau. We get a $D_e$ of $33.02(9)$ Kcal/mol
by VMC, and $32.54(8)$ Kcal/mol by LRDMC, to be compared with the
CCSD(T) value of $32.68$ Kcal/mol from Ref.~\onlinecite{huang}, while
the PBE-DFT gives $D_e=29.55$ Kcal/mol.  The agreement between the
LRDMC and CCSD(T) is impressive, while it is already good for the VMC
estimate.This is mainly due to the size consistency of the JAGP ansatz, 
obtained once the Jastrow is close to the complete basis set limit.
The VMC and CCSD(T) dissociation energies are plotted in 
Fig.~\ref{blowup_landscape} with respect to LRDMC values. 

\subsubsection{Test on a larger molecular cluster}

\begin{table}[!htp]
\begin{ruledtabular}
\caption{Total computational wall time in hours of typical VMC and LRDMC
  runs, performed with the program TurboRVB\cite{sorella_rvb} 
  on 512 thin nodes of the Curie HCP machine (2.7 GHz
  8 core Intel Sandy Bridge processors), to reach a
  target statistical error of $0.06$ Kcal/mol in total energies. Single
  water molecule, protonated dimer and  
a larger cluster of 6 water molecule and one excess proton are
compared. The LRDMC is carried out at a lattice space of 0.125 $a_0$.
}\label{comp_time}
\begin{tabular}{|c|d|d|}
 & \multicolumn{2}{c|}{Total wall time (h) on 512 CPU}  \\
\hline
 \multicolumn{1}{|c|}{$\#$ of water molecules} &
 \multicolumn{1}{c|}{VMC} & \multicolumn{1}{c|}{LRDMC} \\ 
\hline
1 &  0.05 & 0.15 \\
2 & 0.24  & 2.13 \\
6 & 6.49 & 164.35 \\
\end{tabular}
\end{ruledtabular}
\end{table}

QMC methods present a favorable scalability with the number of particles 
with respect to other highly correlated approaches such as CC. 
With the aim at proving this feature also for our approach, we performed 
a benchmark calculation on a more realistic PT model composed of
6 water molecules and one excess proton, which will be the subject of
a further study.
In Table~\ref{comp_time} we report a comparison of the computational
time required to carry out typical VMC and LRDMC runs 
for different sizes of the protonated water cluster at fixed optimized
variational parameters and geometry. The calculations
have been carried out on the HPC Curie thin nodes (2.7 GHz 8 core Intel Sandy Bridge
processors), and
performed with a target statistical error of $0.06$ Kcal/mol in the total energy.
By performing a simple fit on the data in Table~\ref{comp_time}, we notice that
the simple VMC displays a $\sim N^3$ scaling with the number of particles; the
more accurate LRDMC, carried out at a lattice space of $a=0.125 a_0$,
shows instead an almost perfect $N^4$ scaling. 
However, the LRDMC
calculations are still feasible in a reasonable computational time for
the 6 \htwoo\ cluster, while the VMC is still cheap at that cluster size.

\section{Conclusions}
\label{conclusions}

In this paper we presented an extensive study of the protonated water 
dimer by means of the VMC and LRDMC techniques.

The JAGP ansatz employed in this work implements 
an accurate treatment of both static and dynamical correlations 
among electrons. The expansion of the determinantal part over
atomic hybrid orbitals ensures a drastic reduction of the 
number of variational parameters thus making the wave function
optimization procedure efficient and
robust even for large systems.
The comparison with previous published QMC calculations
on the single water molecule showed the quality of our
wave function ansatz. 

Total energy calculations are performed with the less
expensive Variational Monte Carlo approach along with 
the more precise projective Diffusion Monte Carlo. 
The powerful minimization algorithm implemented
in our QMC software, allows an efficient 
estimation of the forces acting on each atomic 
component with a reasonable 
computational cost. Hence both energetics and 
geometry calculations are perfomed within the QMC 
framework. 

A simple mechanism of proton transfer in the dimer 
has been presented and exploited to choose a suitable 
reaction coordinate for the potential energy curve. 
The energy landscape as a function of the oxygen-oxygen 
distance is computed and compared with density functional
theory in the PBE approximation and with CCSD(T) results; LRDMC
is in excellent agreement with CCSD(T) calculations (within $0.2-0.3$
kcal/mol), whereas minor differences (up to about $1$ kcal/mol) are
shown by VMC. Ref.~\onlinecite{alfe} has recently reported 
a similar accuracy for small water clusters by 
diffusion Monte Carlo calculations.

Geometrical properties of the excess proton are also investigated by
VMC structural relaxations, which provide geometries remarkably close
to the ones obtained by a CCSD(T) fitted PES. We show
the presence of two distinct regimes of the dimer
depending on the oxygen-oxygen distance: one with a centrosymmetric 
excess proton and the other with the proton localized on one of the
water molecules. The stability of these 
configurations crucially depends on the level of theory employed to describe the 
electronic structure of the system. 
A better treatment of electron correlation results in the stability
of the asymmetric proton geometry over a wider range of $\overline{OO}$ 
distances. 

These results, together with the  proton transfer static barrier and the 
dissociation energy $D_e$, show that our QMC approach
has a global accuracy comparable with the state-of-the art coupled cluster 
in both geometry and energetics of the dimer, with the advantage of
having a better scaling with the number of particles. 

The accuracy shown by our QMC calculations combines with a favorable
scalability with the number of particles with the respect to other 
highly correlated techniques, as demonstrated with a test
simulation on a larger protonated water cluster. 

These features make this approach a very 
promising candidate for the study of proton transfer in complex acqueous systems. 
In particular, we have shown that the VMC method is cheap, provides
very accurate geometries and a global accuracy of less than 1 kcal/mol
in the most important region for the PT physics.
We hope that our work will inspire further studies on this direction, 
and pave the way for accurate highly-correlated simulations of more 
realistic proton transfer models which will eventually shed new insights 
onto the PT mechanism in water.

\acknowledgments  
We thank Joel M. Bownman and Xinchuan Huang who provided us the coupled cluster
potential energy surface of $H_5O_2^+$. 
One of us (Michele Casula) acknowledges computational resources in the form of the
GENCI grant number x2013096493.


\begin{thebibliography}{99}
 \bibitem{force_field1}  W. L. Jorgensen, J. Chandrasekhar, J. D. Madura, R. W. Impey, M. L. Klein, J. Chem. Phys. \textbf{79}, 926 (1983).
 \bibitem{force_field2} M. Sprik, J. Phys. Chem. \textbf{95},2283 (1991).
 \bibitem{force_field3} M. W. Mahoney, W. L. Jorgensen, J. Chem. Phys. \textbf{112}, 8910 (2000).
 \bibitem{water1_parri} P. L. Silvestrelli, M. Parrinello, J. Chem. Phys \textbf{111}, 3572 (1999).
 \bibitem{water2} J. C. Grossman, E. Schwegler, E. W. Draeger, F. Gygi, G. Galli, J. Chem. Phys. \textbf{120}, 300 (2004).
 \bibitem{water3} J. VandeVondele, F. M. and M. Krack, J. Hutter, M. Sprik, M. Parrinello, J. Chem. Phys. \textbf{122}, 014515 (2005).
 \bibitem{water4} R. Jonchiere, A. P. Seitsonen, G. Ferlat, A. M. Saitta, R. Vuilleumier, J. Chem. Phys. \textbf{135}, 154503 (2011).
 \bibitem{water5} R. Z. Khaliullin, T. D. K\"uhne, arXiv/1303.2067 (2013).
 \bibitem{abinitio_bio1} P. Carloni, U. Rothlisberger, M. Parrinello , Acc. Chem. Res. \textbf{35}, 455 (2002).
 \bibitem{abinitio_bio2} R. A. Friesner, B. D. Dunietza, Acc. Chem. Res. \textbf{34}, 351 (2001).
 \bibitem{water_temperature} S. Yoo, X. C. Zeng, S. S. Xantheas, J. Chem. Phys. \textbf{130}, 221102 (2009).
 \bibitem{water_radial} H. S. Lee, M. E. Tuckerman, J. Chem. Phys. \textbf{126}, 164501 (2007).
 \bibitem{grotthuss} C. J. D. von Grotthuss, Ann. Chim. \textbf{LVIII}, p. 54  (1806).
 \bibitem{agmon} N. Agmon, Chem. Phys. Lett. \textbf{244}, 456 (1995).
 \bibitem{200years_after} D. Marx, ChemPhysChem, \textbf{7}, 1848 (2006).
 \bibitem{phys_rev} T. E. Decoursey, Physiol Rev. \textbf{83}, 475 (2003).
 \bibitem{pnas_membrane} E. Freiera, S. Wolfb, K. Gerwerta, Proc. Natl. Acad. Sci. \textbf{108},11435 (2011).
 \bibitem{photos_ref1} J. Deisenhofer, O. Epp, I. Sinning, H. Michel, J. Mol. Biol. \textbf{246}, 429 (1995).
 \bibitem{photos_ref2} D. Lancaster, H. Michel, B. Honig, M. Gunner, Biophys. J. \textbf{70}, 2469 (1996). 
 \bibitem{PTacqueous_sol1} M. Eigen, Angew. Chem. Int. Ed. \textbf{3}, 1 (1964).
 \bibitem{PTacqueous_sol2} O. F. Mohammed, D. Pines, J. Dreyer, E. Pines, E. T. J. Nibbering, Science \textbf{310}, 83 (2005).
 \bibitem{PTacqueous_sol3} O. F. Mohammed, D. Pines, E. Pines, E. T. J. Nibbering, Chem. Phys. \textbf{341}, 240 (2007).
 \bibitem{luz_exp} Z. Luz, S. Meiboom, J. Am. Chem. Soc. \textbf{86}, 4768 (1964).
 \bibitem{agmon_exp} Noam Agmon, \emph{Hydrogen bonds, water rotation and proton mobility.}, J. Chim. Phys. (Paris), \textbf{93}, 1714 (1996).
 \bibitem{thecomputer} U. W. Schmitt, G. A. Voth, J. Chem. Phys. \textbf{111}, 9361 (1999).
 \bibitem{bond_order} H. Lapid, N. Agmon, M. K. Petersen, G. A. Voth, J. Chem. Phys. \textbf{122}, 014506 (2005).
 \bibitem{parri_nature} D. Marx, M. E. Tuckerman, J. Hutter, M. Parrinello, Nature \textbf{397}, 601 (1999).
 \bibitem{pair_dance} O. Markovitch, H. Chen, S. Izvekov, F. Paesani, G. A. Voth, N. Agmon, J. Phys. Chem. B \textbf{112}, 9456 (2008). 
 \bibitem{chem_rev_pt} D. Marx, A. Chandra, M. E. Tuckerman, Chem. Rev. \textbf{110}, 2174–2216 (2010). 
 \bibitem{marco_prl} A. M. Saitta, F. Saija, P. V. Giaquinta, Phys. Rev. Lett. \textbf{108}, 207801 (2012) 
 \bibitem{hassanali_PT} A. Hassanali, F. Giberti, J. Cuny, T. D. Kühne, M. Parrinello, Proc. Natl. Acad. Sci. USA \textbf{110}, 13723 (2013).
 \bibitem{eigen_model} E. Wicke, M. Eigen, H. Ackermann, Z. Phys. Chem. (N.F.) \textbf{1}, 340 (1954).
 \bibitem{zundel_model} G. Zundel, H. Metzger, Z. Phys. Chem \textbf{58}, 225 (1968).
  \bibitem{extensive_H5O2} Y. Xie, R. B. Remington, H. F. Schaefer, J. Chem. Phys. \textbf{101}, 4878-4884 (1994).
 \bibitem{brueckner} E. F. Valeev, H. F. Schaefer, J. Chem. Phys. \textbf{108}, 7197-7201 (1998).
 \bibitem{wales} D. J. Wales, J. Chem. Phys. \textbf{110}, 10403 (1999).
 \bibitem{aaauer} A. A. Auer, T. Helgaker, and W. Klopper, Phys. Chem. Chem. Phys. \textbf{2}, 2235 (2000).
 \bibitem{DFT_zundel} D. Wei, D. R. Salahub, J. Chem. Phys. \textbf{101}, 7633 (1994).
 \bibitem{parri_science} M. E. Tuckerman, D. Marx, M. L. Klein,M. Parrinello, Science \textbf{275}, 817 (1997).
 \bibitem{zundel_vibrational} J. M. Headrick, E. G. Diken, R. S. Walters, N. I. Hammer, R. A. Christie, J. Cui,
   E. M. Myshakin, M. A. Duncan, M. A. Johnson, K. D. Jordan, Science \textbf{308}, 1765 (2005). 
 \bibitem{huang}X. Huang, B. J. Braams, J. M. Bowman, J. Chem. Phys. \textbf{122}, 044308 (2005).
 \bibitem{espresso} P. Giannozzi \emph{et al.}, J. Phys.: Condens. Matter \textbf{21}, 395502 (2009).
 \bibitem{sorella_rvb} \url{http://people.sissa.it/~sorella/web/}
 \bibitem{pbe} J. P. Perdew, K. Burke, and M. Ernzerhof, Phys. Rev. Lett., \textbf{78}, 1396 (1997). 
 \bibitem{parri_jctc} T. D. Kh\"u ne, M. Krack, M. Parrinello, J. Chem. Theory Comput. \textbf{5}, 235 (2009)
 \bibitem{trouil_mart} Trouiller N., Martins J., Phys. Rev. B, \textbf{43}, 1993 (1991).
 \bibitem{car_parri} R Car, M Parrinello, Phys. Rev. Lett. \textbf{55}, 2471 (1985).
 \bibitem{kato} T. Kato, Comm. Pure Appl. Math. \textbf{10},151 (1957).
 \bibitem{holzman} M. Holzmann, D. M. Ceperley, C. Pierleoni, and K. Esler Phys. Rev. E \textbf{68}, 046707 (2003).
 \bibitem{michele_benzene} S. Sorella, M. Casula, D. Rocca, J. Chem. Phys. \textbf{127}, 014105 (2007).
 \bibitem{azadi} S. Azadi, C. Cavazzoni, S. Sorella, Phys. Rev. B \textbf{82}, 125112 (2010)
 \bibitem{filippi_pseudo} M. Burkatzki, C. Filippi, M. Dolg, J. Chem. Phys. \textbf{126}, 234105 (2007).
 \bibitem{estim1} R. Assaraf and M. Caffarel, J. Chem. Phys. \textbf{119}, 10536 (2003).
 \bibitem{estim2} S. Chiesa, D. Ceperley, S. Zhang, Phys. Rev. Lett. \textbf{94}, 036404 (2005).
 \bibitem{estim3} A. Badinski, P. D. Haynes, J. R. Trail, R. J. Needs,  J. Phys.: Condens. Matter \textbf{22}, 074202 (2010).
 \bibitem{umrigar_warp} C. Filippi, C. J. Umrigar, Phys. Rev. B \textbf{61}, 16291 (2000).
 \bibitem{sorella_warp} S. Sorella, L. Capriotti, J. Chem. Phys. \textbf{133}, 234111 (2010).
 \bibitem{attaccalite_sorella} C. Attaccalite and S. Sorella, Phys. Rev. Lett. \textbf{100}, 114501 (2008).
 \bibitem{pauli} L. Pauling, \emph{The nature of the chemical bond}, 3rd edition, Cornell University Press, Ithaca, New York.
 \bibitem{michele_agp1} D. Nissenbaum, L. Spanu, C. Attaccalite, B. Barbiellini, A. Bansil, Phys. Rev. B \textbf{79}, 035416 (2009).
 \bibitem{sorella_hb} F. Sterpone, L. Spanu, L. Ferraro, S. Sorella, L. Guidoni, J. Chem. Theory Comput., \textbf{4}, 1428 (2008).
 \bibitem{michele_agp2} M. Marchi, S. Azadi, M. Casula, S. Sorella, J. Chem. Phys. \textbf{131}, 154116 (2009).
 \bibitem{zen_water} A. Zen, Y. Luo, S. Sorella, L. Guidoni, J. Chem. Theory Comput. \textbf{9}, 4332 (2013).
 \bibitem{SR1} S. Sorella, L. Capriotti, Phys. Rev. B, \textbf{61}, 2599 (2000).
 \bibitem{SR2} S. Sorella, Phys. Rev. B, \textbf{64}, 024512 (2001).
 \bibitem{SR3} S. Sorella, Phys. Rev. B, \textbf{71}, 241103 (2005).
 \bibitem{SR4} C. J. Umrigar, J. Toulouse, C. Filippi, S. Sorella, R. G. Hennig, Phys. Rev. Lett. \textbf{98}, 110201 (2007).
 \bibitem{michele_lrdmc} M. Casula, C. Filippi, S. Sorella, Phys. Rev. Lett. \textbf{95}, 100201 (2005).
 \bibitem{filippi_lrdmc} M. Casula, S. Moroni, S. Sorella, C. Filippi, J. Chem. Phys. \textbf{132}, 154113 (2010).
 \bibitem{exp_H2O_geometry} W. S. Benedict, N. Gailar, E. K. Plyler,
   J. Chem. Phys. \textbf{24}, 1139 (1956).
\bibitem{fink} A. L\"uchow and R. Fink, J. Chem. Phys. \textbf{113},
  8457 (2000).
\bibitem{exact_H2O_energy} D. Feller, C. Boyle, and E. Davidson, J. Chem. Phys. \textbf{86}, 3424 (1987).
 \bibitem{CCSD_zundel} M. Park, I. Shin, N. J. Singh, K. S. Kim, J. Phys. Chem. A, \textbf{111}, 10692-10702 (2007).
 \bibitem{water_dist} U. Bergmann, A. Di Cicco, P. Wernet, E. Principi, P. Glatzel et al, J. Chem. Phys. \textbf{127}, 174504 (2007).
 \bibitem{parri_science_ref} S. S. Xantheas, J. Am. Chem. Soc. \textbf{117}, 10373 (1995).
 \bibitem{pavese} M. Pavese, S. Chawla, D. Lu, J. Lobaugh, G.A. Voth,  J. Chem. Phys. \textbf{107}, 7428 (1997).
 \bibitem{adamo} S. Sadhukhan, D. Munoz, C. Adamo, G. E. Scuseria, Chem. Phys. Lett \textbf{306}, 83-87 (1999).
 \bibitem{eric_size_cons} E. Neuscamman, Phys. Rev. Lett. \textbf{109}, 203001 (2012)
 \bibitem{alfe} M. J. Gillan, F. R. Manby, M. D. Towler, D. Alf\`e, J. Chem. Phys. \textbf{136}, 244105 (2012)  
\end{thebibliography}
\end{document}